\documentstyle[aps,multicol,epsf]{revtex}
\def\narrowcaption{\columnwidth20.5pc}

\begin{document}

\title{Josephson-Junction Qubits and the 
Readout Process by Single-Electron Transistors}

\author {Yuriy Makhlin$^{1,2}$, Gerd Sch\"on$^{1}$, 
         and Alexander Shnirman$^{1,3}$}

\address{
$^1$Institut f\"ur Theoretische Festk\"orperphysik,
Universit\"at Karlsruhe, D-76128 Karlsruhe, Germany. \\
$^2$Landau Institute for Theoretical Physics, 117940 Moscow, Russia.\\
$^3$Department of Physics, University of Illinois at Urbana-Champaign,
Urbana, IL 61801-3080, U.S.A.}

\maketitle
\def\thefootnote{\fnsymbol{footnote}}

\begin{abstract}
Several physical realizations of quantum bits have been proposed, 
including trapped ions, NMR systems and spins in nano-structures, 
quantum optical systems, and nano-electronic devices.  The latter 
appear most suitable for large-scale integration and potential 
applications.  We suggest to use low-capacitance Josephson junctions, 
exploiting the coherence of tunneling in the superconducting state 
combined with the possibility to control individual charges by 
Coulomb blockade effects. These systems constitute quantum bits 
(qubits), with logical states differing by one Cooper-pair charge. 
Single- and two-bit operations can be performed by applying a sequence 
of gate voltages. The phase coherence time is sufficiently long to 
allow a series of these steps. The ease and precision of these 
manipulations depends on the specific design. Here we concentrate 
on a circuit which is most easily fabricated in an experiment.

In addition to the manipulation of qubits the resulting quantum state has to 
be read out.  This can be accomplished by coupling a single-electron transistor
capacitively to the qubit.  To describe this quantum measurement process we
study the time evolution of the density matrix of the coupled system.  Only 
when
a transport voltage is turned on, the transistor destroys the phase coherence 
of
the qubit; in this case within a short time.  The measurement is accomplished
after a longer time, when the signal resolves the different quantum
states.  At still longer times the measurement process itself destroys the
information about the initial state.  We present a suitable set of system
parameters, which can be realized by present-day technology.
\end{abstract}

\begin{multicols}{2}
\section{Introduction}

The investigation of nano-scale electronic devices, such as low-capacitance
tunnel junctions or quantum dot systems, has always been motivated by the
perspective of future applications.  By now several have been demonstrated,
e.g.\ the use of single-electron transistors (SET) as ultra-sensitive
electro-meters and single-electron pumps.  From the beginning it also appeared
attractive to use these systems for digital operations needed in classical
computation (see Ref.~\onlinecite{Likharev} for a review).  Obviously 
{\sl single-electron} devices would constitute the ultimate electronic memory.
Unfortunately, their extreme sensitivity makes them also very susceptible to
fluctuations and random background charges.  Due to these problems --- and the
continuing progress of conventional techniques --- the future of SET devices in
{\sl classical} digital applications remains uncertain.

The situation is different when we turn to elements for quantum computers.  
They
could perform certain calculations which no classical computer could do in
acceptable times by exploiting the quantum mechanical coherent evolution of
superpositions of states~\cite{Barenco_Review}.  Here conventional systems
provide no alternative.  In this context, ions in a trap, manipulated by laser
irradiation are the best studied system \cite{Zoller,Wineland}.  However,
alternatives need to be explored, in particular those which are more easily
embedded in an electronic circuit.  From this point of view nano-electronic
devices appear particularly promising.

The simplest choice, normal-metal single-electron devices are ruled out, since
--- due to the large number of electron states involved --- different,
sequential tunneling processes are incoherent.  Ultra-small quantum dots with
discrete levels, and in particular, spin degrees of freedom embedded in
nano-scale structured materials are candidates. They can be manipulated by 
tuning
potentials and barriers~\cite{Loss}.  However, these systems are difficult to
fabricate in a controlled way.  More attractive appear systems of Josephson
contacts, where the coherence of the superconducting state can be exploited and
the technology is quite advanced. Macroscopic quantum effects associated with
the flux in a SQUID have been demonstrated~\cite{Lukens}.  Quantum extension of
elements based on single flux logic have been suggested~\cite{Mooij,Ioffe}, and
efforts are made to observe one of the elementary processes, the coherent
oscillation of the flux between degenerate states~\cite{Rome}.

We have suggested~\cite{Our_PRL} to use low-capacitance Josephson junctions,
where Cooper pairs tunnel coherently while Coulomb blockade effects allow the
control of individual charges. They provide physical realizations of quantum
bits (qubits) with logical states differing by the number of Cooper pair 
charges
on an island.  These junctions can be fabricated by present-day technology. 
The coherent tunneling of Cooper pairs and the related properties of quantum
mechanical superpositions of charge states have been discussed and demonstrated
in experiments~\cite{Bouchiat_PhD,Geerligs,Parity,Siewert,Nakamura,Hadley}.
In particular, in a recent experiment~\cite{Nakamura_Nature} Nakamura et~al. 
have observed time-resolved coherent oscillations of charge in such Josephson 
junction setup.  

We concentrate here on the simplest design of the qubit, which is sufficient 
to introduce the ideas, and it is most easy for fabrication. It should be 
added that other ideas in this direction have been 
discussed~\cite{Averin,Makhlin}. 
In Ref.~\cite{Makhlin} an improved design has been proposed that relaxes 
requirements to the parameters of the circuit and makes the manipulation 
procedures more flexible. In Section~\ref{Section_Qubits} we will introduce the 
system and show how single- and two-bit operations (gates) can be performed 
by the application of sequences of gate voltages. In 
Section~\ref{Section_Dephasing} 
we analyze the influence of dissipation and fluctuations and conclude that for 
a proper choice of parameters the phase coherence time is sufficiently long to 
allow a large number of gate operations.

In addition to the controlled manipulations of qubits, quantum computation
requires a quantum measurement process to read out the final state.  The
requirements for both steps appear to contradict each other.  During
manipulations the dephasing should be minimized, whereas a measurement should
dephase the state of the qubit as fast as possible.  The option to couple the
measuring device to the qubit only when needed is hard to achieve in nano-scale
systems.  The alternative described in Section~\ref{Section_Measurement}, is to
couple a normal-state single-electron transistor capacitively to a
qubit~\cite{Shnirman_PRB}.  During the manipulations the transport voltage of
the SET is turned off, and the SET only acts as an extra capacitor.  To perform
the measurement the transport voltage is turned on.  The dissipative current
through the transistor dephases the qubit and provides the read-out signal for
the quantum state.  We describe this quantum measurement process by considering
explicitly the time-evolution of the density matrix of the coupled system.  We
find that the process is characterized by three time scales:  a short dephasing
time, the longer `measurement time' when the signal resolves the different
quantum states, and finally the mixing time after which the measurement process
itself destroys the information about the initial state.  Similar 
nonequilibrium
dephasing processes~\cite{Levinson,Aleiner,Gurvitz} have recently been
demonstrated experimentally~\cite{Buks}.

In Section~\ref{Section_Discussion} we discuss system parameters and suggest
suitable sets which can be realized by present-day technology.  We further
compare with related work.


\section{Josephson junction qubits}
\label{Section_Qubits}

\subsection{Qubits and single-bit gates}   

\begin{figure}
\epsfysize=14\baselineskip
\centerline{\hbox{\epsffile{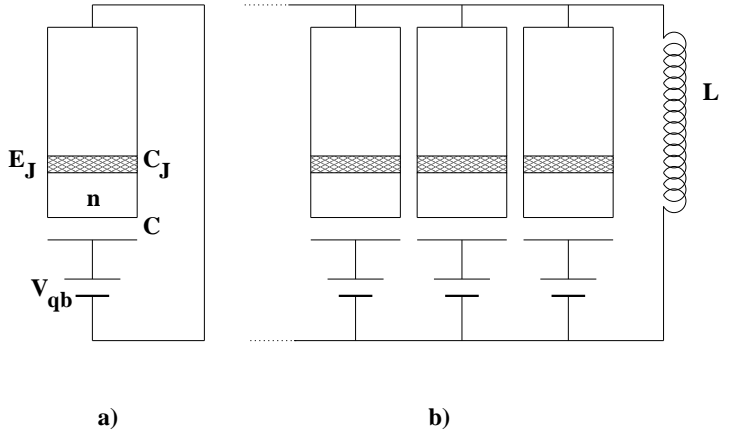}}}
\vskip 0.8cm
\narrowcaption
\caption[]{\label{BITS_IDEAL}
Ideal a) one qubit and b) multi-qubit systems.}
\end{figure}

The simplest Josephson junction qubit is shown in Fig.~\ref{BITS_IDEAL}a.  It
consists of a small superconducting island connected by a tunnel junction, with
capacitance $C_{\rm J}$ and Josephson coupling energy $E_{\rm J}$, to a
superconducting electrode.  An ideal voltage source, $V_{\rm qb}$, is connected
to the system via a gate capacitor $C$ (fluctuation effects will be discussed
later).  We choose a material such that the superconducting energy gap $\Delta$
is the largest energy in the problem, larger even than the single-electron
charging energy to be discussed below.  In this case quasi-particle tunneling 
is
suppressed at low temperatures, and a situation can be achieved where no
quasiparticle excitations exist on the island\footnote{Under suitable
conditions the superconducting state is totally paired, i.e.\ the number of
electrons on the island is even, since an extra quasi-particle (odd number of
electrons) costs the extra energy $\Delta$.  This `parity effect' has been
established in experiments below a crossover temperature $T^{*} \approx \Delta
/\ln N_{\rm eff}$, where $N_{\rm eff}$ is the number of electrons in the system
near the Fermi energy~\cite{Parity,Lafarge,Schoen-Zaikin94,Tinkham2}.  For a
small enough island, $T^{*}$ is typically one order of magnitude smaller than
the superconducting transition temperature.  For the case that --- e.g.\ due to
the initial preparation --- an unpaired excitation exists on the island, a
channel should be provided for the quasiparticle to escape to normal parts of
the system~\cite{Lafarge}.}.

In the following we will consider the situation
where only Cooper pairs tunnel in the superconducting junction.   
This system is described by  the Hamiltonian
\begin{equation}
\label{1bit_Hamiltonian_Eqb}
        H = 4E_{\rm qb}(n - n_{\rm qb})^2 - 
        E_{\rm J} \cos\Theta \ . 
\end{equation}
Here, $n$ is the number (operator) of extra Cooper pair charges 
on the island (relative to some neutral reference state)
and the phase variable $\Theta$ is its
conjugate $n = -i\hbar\,\partial/\partial(\hbar \Theta)$. 
The charging energy of the superconducting island is characterized by the scale
$E_{\rm qb} \equiv e^2/2(C +  C_{\rm J})$,
while the dimensionless gate charge,
$n_{\rm qb} \equiv CV_{\rm qb}/2e$, acts as a control field.
We consider systems where  $E_{\rm qb}$ 
is much larger than the Josephson coupling energy,
$E_{\rm qb} \gg E_{\rm J}$.
In this regime a convenient basis is formed by the charge states, 
 parameterized by the number of Cooper pairs $n$ on the island. 
In this basis the Hamiltonian (\ref{1bit_Hamiltonian_Eqb}) reads
\begin{eqnarray}
H =&& \sum_n 4E_{\rm qb}(n - n_{\rm qb})^2 |n\rangle \langle n|
\nonumber\\
&& - \frac{1}{2}E_{\rm J} 
\Big(|n\rangle \langle n+1| + |n+1\rangle \langle n|\Big) \; .
\end{eqnarray}

\begin{figure}
\epsfysize=10\baselineskip
\centerline{\hbox{\epsffile{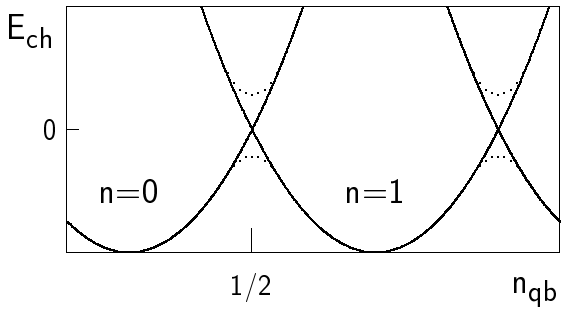}}}
\vskip 0.8cm
\narrowcaption
\caption[]{\label{Parabolas}
The charging energy of the superconducting electron box 
is shown (solid lines) as a function of 
the gate charge $n_{\rm qb}$ for different 
numbers of extra Cooper pairs $n$ on the island. 
Near degeneracy points the weaker Josephson coupling energy 
mixes the charge states and modifies the energy of the 
eigenstates (dotted line). In this regime the system 
effectively reduces to a two-state quantum system.
}
\end{figure}

For most values of $n_{\rm qb}$ the energy levels are dominated by the
charging part of the Hamiltonian.
However, when
$n_{\rm qb}$ is approximately half-integer and the charging energies of two 
adjacent states are close to each other,
the Josephson tunneling mixes them strongly (see Fig.~\ref{Parabolas}).

We concentrate on the voltage interval near a degeneracy point
of two charge states, say $n=0$ and $n=1$. 
For the parameters chosen, further charge states can be ignored, and
the system (\ref{1bit_Hamiltonian_Eqb}) reduces to a two-state model,
with Hamiltonian which can be written in spin-$1\over2$ notation
in terms of Pauli spin matrices 
$\vec{\sigma} =  \sigma_x,\sigma_y,\sigma_z$ as
\begin{equation}
\label{Magnetic_Hamiltonian}
	H= \frac{1}{2}E_{\rm ch}(n_{\rm qb}) \sigma_z - 
	{1\over 2} E_{\rm J} \sigma_x \ . 
\end{equation}
The charge states $n=0$ and  $n=1$ correspond to the spin 
basis states $\left|\downarrow \right.\rangle \equiv \left(_1^0
\right)$ and $\left| \uparrow\right.\rangle  \equiv \left( _0^1 \right)$,
respectively. The charging energy $E_{\rm ch}(n_{\rm qb})
=4E_{\rm qb}(1-2n_{\rm qb})$, equivalent to a magnetic field in 
$z$-direction in the spin problem, is controlled by the gate voltage. For 
convenience we can further rewrite the Hamiltonian as
\begin{equation}
\label{Magnetic_Hamiltonian'}
	H=\frac{1}{2}\Delta E(\eta) 
	\left(\cos\eta\ \sigma_z-\sin\eta\ \sigma_x \right) \; ,
\end{equation}
where the mixing angle 
$\eta=\tan^{-1}[E_{\rm J}/E_{\rm ch}(n_{\rm qb})]$ determines the 
direction of the effective magnetic field in the $xz$-plane, and the energy 
splitting between the eigenstates is $\Delta E(\eta) = 
\sqrt{E_{\rm ch}^2(n_{\rm qb}) + E_{\rm J}^2} = E_{\rm J}/\sin\eta$. 
At the degeneracy point, $\eta = \pi/2$, it reduces to $E_{\rm J}$.
The eigenstates  are denoted in the following as $|0\rangle$  and $|1\rangle$. 
For some chosen value of  $n_{\rm qb}$ they are
\begin{eqnarray}
\label{Eigen_Basis}
	|0\rangle  &=& \;\;\;  \cos{\eta\over2} \left| \downarrow 
\right.\rangle 
+ 
	\sin{\eta\over2} \left| \uparrow \right.\rangle \nonumber \\ 
	|1\rangle &=& -  \sin{\eta\over2} \left| \downarrow \right.\rangle +
	\cos{\eta\over2} \left| \uparrow \right.\rangle \ .
\end{eqnarray}

For later convenience we can rewrite the Hamiltonian in the basis of
eigenstates.  To avoid confusion we introduce a second set of Pauli matrices,
$\vec{\rho}$, which operate in the basis $|0\rangle$ and $|1\rangle$, while
reserving $\vec{\sigma}$ for the basis of charge states $\left| \downarrow
\right.\rangle$ and $\left| \uparrow \right.\rangle$.  By definition the
Hamiltonian then becomes $H=\frac{1}{2}\Delta E(\eta)\rho_z$.

By changing the gate voltage we can perform the required 
one-bit operations (gates). If, for example, one chooses 
the idle state far away from degeneracy, the eigenstates 
$|0\rangle$ and $|1\rangle$ are close to $\left| \downarrow \right.\rangle$ 
and $\left| \uparrow \right.\rangle$, respectively.
For definiteness, we choose the eigenstates $|0\rangle$ and $|1\rangle$ at the 
idle point as logical basis states.
Then switching the system suddenly to the degeneracy point for a time
$\Delta t$ and then switching back produces a rotation in spin space,
\begin{equation}
\label{1bit_spin_flip}
U_{\rm 1-bit}(\Delta t)=  \exp \big(i \alpha \sigma_x \big)
= \left( 
\begin{array}{cc}
\cos\alpha  & i\sin\alpha \\ 
i\sin \alpha & \cos\alpha
\end{array} 
\right) 
\; ,
\end{equation}
where $\alpha = E_{\rm J} \Delta t/ 2 \hbar$.
Depending on the value of $\Delta t$, a spin flip can be produced, 
or, starting from $|0\rangle$,  a superposition of states with any 
chosen weights can be reached. This is exactly the way the experiments
of Nakamura~et~al. \cite{Nakamura_Nature} were performed.
At the same time, keeping the system at the idle point we achieve a phase shift 
between two logical states by the angle $\Delta E(\eta_{\rm idle})\Delta t$.
A different phase shift during the same time period $\Delta t$ can be achieved 
using a temporary change of $n_{\rm qb}$ by a small amount which changes the
energy difference between the ground and excited states.

The example presented above provides an approximate spin flip 
for the situation where the idle point is far from degeneracy and 
$E_{\rm qb} \gg E_{\rm J}$. But a spin flip in the logical basis can also be 
performed exactly. 
It requires that we switch from the idle point $\eta_{\rm idle}$ to the 
point where the effective magnetic field is orthogonal to the idle one, 
$\eta=\eta_{\rm idle} + \pi/2$. This changes the Hamiltonian from 
$H=\frac{1}{2}\Delta E(\eta_{\rm idle})\rho_z$ to 
$H=\frac{1}{2}\Delta E(\eta_{\rm idle}+\pi/2)\rho_x$. 
To achieve that, the dimensionless gate charge $n_{\rm qb}$ 
should be increased by $E_{\rm J}/(4E_{\rm qb}\sin 2\eta_{\rm idle})$. 
In the limit discussed above, $\eta_{\rm idle} \ll 1$, the operating 
point is close to the degeneracy, $\eta=\pi/2$.

Unitary $\rho_x$- and $\rho_z$-rotations described above are sufficient for all
manipulations with a single qubit.  By using a sequence (of no more than three)
such elementary rotations we can achieve any unitary transformation of the
qubit's state.

In our discussion so far elementary rotations should have been performed one 
immediately after another. However, sometimes the quantum state should be kept 
intact for a certain time interval, for instance, while other qubits are 
manipulated during the computation.
Even in the idle state $\eta=\eta_{\rm idle}$, 
the energies of the two eigenstates are different.
Hence their phases evolve relative to each other, which leads  
to the quantum mechanical `coherent oscillations' of a system 
which is in a superposition of eigenstates. We have to keep 
track of this time dependence as is demonstrated by the following example
(for simplicity again on the approximate level).
Imagine we try to perform a rotation by angle $\alpha$ in the spin space,
which we can do by a suitable choice of $\Delta t$ in
(\ref{1bit_spin_flip}), and after some time delay $\tau$ we perform the 
reverse operation. The unitary transformation for this combined process is
$$
\exp \big(-i \alpha \sigma_x \big) \cdot 
\exp \left(i \frac{\Delta E(\eta_{\rm idle})\tau}{2\hbar} \sigma_z \right)
\cdot \exp \big(i \alpha \sigma_x \big) \; .
$$
Clearly the result depends on the intermediate time $\tau$
and differs from unity or a simple phase shift, 
unless $\Delta E(\eta_{\rm idle})\tau/2\hbar = n \pi$.
The time-dependent phase factors, arising from
the energy difference in the idle state, can be removed 
from the eigenstates if all the calculations are performed
in the interaction representation, with zero-order Hamiltonian 
being the one at the idle point. In this way the information which
is contained in the amplitudes of the qubit's states is 
preserved. However, there is a price
for this simplification, namely the transformation to 
the interaction representation 
introduces additional time dependence in the Hamiltonian during 
the operations. 
Thus a general 1-bit operation induced by
switching at  $t_0$ from $\eta_{\rm idle}$ to $\eta$ for some time
$\Delta t$ is  described by the unitary transformation 
\begin{eqnarray}
{\cal U}(t_0,\Delta t,\eta) =
e^{{i\over \hbar}H(\eta_{\rm idle})(t_0+\Delta t)}
\, e^{-{i\over \hbar}H(\eta)\Delta t}
\, e^{-{i\over \hbar}H(\eta_{\rm idle})t_0} \, .
\label{Ugeneral}
\end{eqnarray}
This demonstrates that  the effect of 
the operations depends not only on their time span  but also on 
the moment when they start.

The transformation (\ref{Ugeneral}) is the elementary operation in the
interaction picture in the sense that it can be performed in one step (one
voltage switching).  Can we perform an arbitrary single-bit gate using such an
operation?  To answer this question we note that all the unitary 
transformations
of the two-state system form the group SU(2).  The group is three-dimensional,
hence one needs in general three controllable parameters to obtain a given
single-bit operation.  As is obvious from (\ref{Ugeneral}), $\cal U$ depends
exactly on three parameters:  $\eta$, $t_0$ and $\Delta t$. It can
be shown that {\sl any} single-bit gate can be performed in one step if the
voltage (i.e.\ $\eta$) and the times $t_0$, $\Delta t$ are chosen
properly.\footnote{It is customary to consider rotations about two axes, say,
$x$- and $z$-rotations, as device-independent elementary single-qubit gates.
Then, any single-bit gate in a quantum algorithm is described as a combination
of three consecutive elementary rotations given by their angles (e.g.\ Euler
angles $\alpha$, $\beta$, $\gamma$).  To realize such a gate $\cal O$ we need 
to
solve the equation ${\cal U}(\eta,t_0,\Delta t)={\cal O}(\alpha,\beta,\gamma)$
to find the triple of $\eta$, $t_0$ and $\Delta t$.  Since usually only a few
different single-bit gates are used, corresponding triples can be calculated in
advance.}  (The dependence on $t_0$ is periodic with the period $h/\Delta
E(\eta_{\rm idle})$.  Hence the waiting time to a new operation is restricted
to this period.)

As a final remark in this subsection, we notice that our choice of the logical 
basis is by no means unique. As follows from the preceding discussion, we can 
perform $x$- and $z$-rotations in this basis, which provides sufficient tools 
for any unitary operation. On the other hand, since we can perform {\it any} 
unitary transformation, we can as well perform $x$- and $z$-rotations in any 
basis of our two-dimensional system. Therefore, any basis can serve as the 
logical one. The Hamiltonian at the idle point is diagonal in the basis 
(\ref{Eigen_Basis}), while the controlled part of the Hamiltonian, the charging 
energy favors the charge basis. For this reason manipulation procedures for 
$x$- 
and $z$-rotations, in the interaction representation, are of comparable 
simplicity in all bases. The preparation procedure (thermal relaxation at the 
idle point) is easier described in the eigenbasis, while coupling to the meter 
(see Section~\ref{Section_Measurement}) is diagonal in the charge basis. So, 
there is no obvious choice for the logical states and it is a matter of 
convention.

\subsection{Many-qubit system}
\label{Many_Qubit}

For quantum computation a register, consisting of a (large)
number of qubits is needed, and pairs of qubits have to be coupled in
a controlled way.
Such two-bit operations (gates) are necessary, for
instance, to create entangled states. 
For this purpose we couple all qubits by one mutual inductor 
as shown in Fig.~\ref{BITS_IDEAL}b.
One can easily see that for $L=0$ the system reduces to a series of
uncoupled qubits, while for $L=\infty$ they are coupled 
strongly. Finite values of $L$ introduce some retardation 
in the interaction.
The Hamiltonian of this system, consisting of $N$ qubits
and an oscillator formed by the inductance and the total capacitance of
all qubits is~\cite{Our_PRL}
\begin{eqnarray}
	   H = & \sum_{i=1}^N \Big\{4E_{\rm qb} (n_i - n_{\rm qb,i})^2  
\nonumber\\ 
 &      - E_{\rm J} \cos\left(\Theta_i -
		2\pi {C_{\rm t}\over C_{\rm J}} {\Phi\over \Phi_0}\right) 
                \Big\}+ {q^2\over 2NC_{\rm t}} + {\Phi^2\over 2L} \ .
\label{2bit_Hamiltonian_Cosine}
\end{eqnarray}
Here $\Phi$ is the flux in the mutual inductor, 
$\Phi_0 \equiv h/2e$  is the flux quantum, and 
$q = -i\hbar\,\partial/\partial \Phi$, the variable 
conjugated to $\Phi$, is related to 
the total charge on the gate capacitors of the qubits. 
For the oscillations in this $LC$-circuit
the junction and gate capacitor of each qubit act in series, 
hence the relevant capacitance is 
$C_{\rm t}^{-1}= C_{\rm J}^{-1} + C^{-1}$.
The voltage oscillations of the $LC$-circuit affect all the 
qubits equally, thus $\Phi$ is coupled to each 
of the phases $\Theta_i$.
The reduction factor $C_{\rm t}/C_{\rm J}$ describes the screening 
of these voltage oscillations by the gate capacitors.
Here we have chosen to express the coupling via a shift of the 
phase in the Josephson coupling terms arising from the voltage 
oscillations. This form is reminiscent of the usual description
of SQUIDs, except that $\Phi$  is a dynamic variable
and its effect is reduced by the ratio of capacitances. 
Alternatively, we could have added the oscillating charges to the
gate charges in the charging energy. Both forms are 
equivalent and related by a canonical transformation.
This and a more detailed derivation of (\ref{2bit_Hamiltonian_Cosine})
has been presented in Ref.~\onlinecite{Our_PRL}.

The oscillator described by the charge $q$ and flux $\Phi$ 
with characteristic frequency
$\omega_{LC}^{(N)} = (NLC_{\rm t})^{-1/2}$ produces 
an effective coupling between the qubits. We choose parameters such that 
\begin{eqnarray}
\label{High_Frequency_Condition}
\Delta E(\eta) & \ll & \hbar\omega_{LC}^{(N)} \ ,\\
\label{Small_Fluctuation_Assumption}
(C_{\rm t}/C_{\rm J})\sqrt{\langle\Phi^2\rangle} 
& \ll &   \Phi_0 \ .
\end{eqnarray}
The first condition (\ref{High_Frequency_Condition}) assures 
that the oscillator remains in its ground state at all relevant
operation frequencies. I.e.\ the logical operations on the qubits
are not affected by excited states.
The second condition (\ref{Small_Fluctuation_Assumption})
prevents the Josephson coupling in (\ref{2bit_Hamiltonian_Cosine})
from being ``washed out'' by the fluctuations of $\Phi$.
Since they are limited by $L$, hence
 $\langle\Phi^2\rangle / L \approx {1\over2} \hbar  \omega_{LC}^{(N)}$,
this condition imposes only weak constraints on the parameters.

Although the $LC$-oscillator remains in its ground state, it provides an
effective coupling between the qubits. To analyze this we expand in
(\ref{2bit_Hamiltonian_Cosine}) the Josephson coupling terms in $\Phi$
and, because of (\ref{Small_Fluctuation_Assumption}), neglect powers
higher than linear. The linear term is $I\Phi$, where the
current through the inductor is given by the sum of qubits' contributions,
\begin{equation}
I = {C_{\rm t}\over C_{\rm J}} {2\pi E_{\rm J}\over \Phi_0}
\sum_i \sin\Theta_{i}\ .
\end{equation}
The linear term and the unperturbed Hamiltonian of
the oscillator can be combined to a square,
\begin{equation}
\frac{q^2}{2NC_t} + \frac{\Phi^2}{2L} +I\Phi=
\frac{q^2}{2NC_t} + \frac{(\Phi+L I)^2}{2L} -\frac{LI^2}{2}
\label{PerturbedHam}\ .
\end{equation}
As long as the frequency of the oscillator is large compared to
characteristic frequencies of the qubit's motion
(\ref{High_Frequency_Condition}), one can use the adiabatic
approximation and treat the slow qubit's variables ($\Theta_i$) as
constant to find the energy levels of the oscillator. Since the lowest
level of the first two terms in the rhs of Eq.(\ref{PerturbedHam}),
equal to $\hbar\omega_{LC}/2$, does not depend on $I$, the correction to
the ground state energy is given by the last term. This term provides
the effective coupling between the qubits
\begin{equation}
H_{\rm int}=-E_L\left(\sum_i \sin\Theta_i\right)^2\ ,
\label{2bit_Coupling}
\end{equation}
where the energy scale is
\begin{equation}
\label{E_L}
E_L = 2\pi^2 {C_{\rm t}^2\over C_{\rm J}^2} {E_{\rm J}^2 L\over\Phi_0^2}\ .
\end{equation}
In the spin-$1\over2$ notation $\sin\Theta_i = \frac{1}{2}\sigma_y^{(i)}$ and
the interaction term becomes (up to constant terms)
\begin{equation}
\label{2bit_Spin_Coupling}
	H_{\rm int} 
	 = - {E_L\over 2} \sum_{i<j} \sigma_y^{(i)}  \sigma_y^{(j)}\ . 
\end{equation}

The ideal system would be one where the coupling between different qubits could
be switched on and off, leaving the qubits uncoupled in the idle state and
during 1-bit operations.  This option requires a more complicated
design~\cite{Makhlin}.  With the present, simplest model the qubits are coupled
at all times.  But even in this case we can control the coupling in an
approximate way by tuning the energies of the selected qubits in and out of
resonance.  If $E_L$ is smaller or comparable to $E_{\rm J}$ and the gates
voltages $V_{{\rm qb},i}$ are all different, such that no two qubits are
near a degeneracy, the interaction (\ref{2bit_Spin_Coupling}) has only weak
effects.  In this case, the eigenstates of, say, a two-qubit system are
approximately $\left| \downarrow \downarrow \right.\rangle$, $\left| \downarrow
\uparrow \right.\rangle$, $\left| \uparrow \downarrow \right.\rangle$ and
$\left| \uparrow\uparrow \right.\rangle$, separated by energies larger
than $E_L$. Hence, the effect of the coupling is weak. If, however, a
pair of these states is degenerate, even a weak coupling lifts the degeneracy, 
changing
the eigenstates drastically.  For example, if $V_1 = V_2$, the states $\left|
\downarrow\uparrow \right.\rangle$ and $\left| \uparrow\downarrow
\right.\rangle$ are degenerate.  In this case the correct eigenstates are:
${1\over\sqrt{2}}(\left| \downarrow\uparrow \right.\rangle + \left|
\uparrow\downarrow \right.\rangle)$ and ${1\over\sqrt{2}}(\left|
\downarrow\uparrow \right.\rangle - \left| \uparrow\downarrow \right.\rangle)$
with energy splitting $E_L$ between them.

With the coupling  (\ref{2bit_Spin_Coupling})
we are able to perform two-bit operations and create, e.g.,
entangled states. In the idle state we bias all  qubits at different
voltages. Then we suddenly switch the voltages of two selected qubits to be 
equal, bringing those two qubits for a time $\Delta t$ into resonance, and then
we switch back. The result is a two-bit operation, which is a rotation in the 
subspace spanned by $\left| \downarrow\uparrow \right.\rangle, \left| 
\uparrow\downarrow \right.\rangle$:
\begin{equation}
\label{2bit_Operation}
U_{\rm 2-bit}(\Delta t)=  
\left( 
\begin{array}{cc}
\cos \beta  & i\sin \beta\\ 
i\sin \beta & \cos \beta
\end{array} 
\right) \ ,
\end{equation}
where $\beta = E_L \Delta t/2 \hbar$. The states $\left| \downarrow\downarrow 
\right.\rangle, \left| \uparrow\uparrow \right.\rangle$ merely acquire phase 
factors, relative to what they would acquire in the idle state. For all other 
qubits the interaction remains a small perturbation.

Similar to the situation with single-bit gates, the many-qubit states
have time-dependent phase factors and the result of consecutive 2-bit
gates depends on the waiting time between them.  Again a transformation to the
interaction representation makes this time dependence explicit. One can show 
that any two-qubit operation can, in principle, be performed exactly.
However, the complexity of such exact manipulation grows with the number 
of qubits. The improved design proposed in~\cite{Makhlin} allows one to switch
on and off the coupling between the qubits and to perform exact two-qubit 
operations in a simple way.  

The two-bit gates (\ref{2bit_Operation}) together with all one-bit gates, 
i.e.\ spin rotations (\ref{1bit_spin_flip}) and the phase shifts, constitute 
a {\sl universal set}:  They are sufficient for all manipulations required
for quantum computation.


\subsection{Extensions and discussion}

To check the experimental feasibility of the present proposal it is
necessary to estimate the time-span $\Delta t$ of the voltage pulses
needed for typical single-bit operations. We note that a reasonable 
value of $E_{\rm J}$ is of the order of $E_{\rm J}/k_{\rm B} 
=0.1$--1K.
It cannot be chosen much lower,
since the condition $k_{\rm B} T \le E_{\rm J}$ should be satisfied,
and it should not be much larger since this would increase the
technical difficulties associated with the time control.  
The corresponding time-span is nevertheless short, 
$\Delta t \approx  \hbar/E_{\rm J} =10$--$100$ps, 
and is difficult to control in an experiment.

There exist, however, alternatives.
On one hand, a controlled ramping of the coupling energy provides a well
defined, though non-trivial single-bit operation.
Since in general, any 1-bit gate can be performed with proper 
choice of 3 controlled parameters, a universal set of gates can be
produced in this way.
Another possibility is to follow the established  procedures of
spin resonance experiments. 
E.g.\ a coherent spin rotation can be performed as follows:
The system is moved adiabatically to the degeneracy point. Then an
ac voltage with frequency $E_{\rm J}/\hbar$ is applied. 
Finally, the system is moved adiabatically back to the idle point. 
The time-width of the ac-pulse needed, e.g.,
for a total spin flip depends on the ac-amplitude, therefore it can be 
chosen much longer then $ \hbar/E_{\rm J}$. Unfortunately,
in comparison to the sudden switching, the slow 
adiabatic approach allows only a reduced number of operations during the
phase coherence time.

Also the two-bit gates, instead of application of short voltage pulses, 
can be performed by moving the system 
adiabatically to a degeneracy point (say $V_1=V_2$), and then applying an ac 
voltage pulse in the antisymmetric channel  
$(V_1 - V_2) \propto \exp(i E_L t)$. 

To improve the performance of the device
generalizations of the design of the qubits and their coupling
may be useful. We mention here  one
which has been described in  Ref.~\onlinecite{Makhlin}. There  
we discuss a system where the Josephson coupling can be tuned to zero as 
well. This can be achieved by replacing each Josephson junction by a 
SQUID with two junctions, and controlling the effective
coupling by an applied magnetic flux. While this design is
more complicated to fabricate, it has considerable advantages: 
(i)  In the 
idle state the Hamiltonian can be tuned to zero ($E_{\rm ch}(V_{\rm qb}) 
= E_{\rm J,eff} = 0$), which removes the
problem that the unitary transformations not only depend on the time
span of their duration but also on the time $t_0$ when they are performed.
(ii) The two-bit coupling is turned on only for those two qubits
which have both  $E_{\rm J,eff} \ne 0$. 
(iii) With only $E_{\rm ch}(V_{\rm qb})$ or $E_{\rm J,eff}$ non-zero, the
unitary transformations depend in a simple way on the time integral 
of the corresponding Hamiltonian. Hence ramping the energies produces
simple, well-defined gates.
As a result of these improvements the 
manipulations can be performed with much higher accuracy.


\section{Circuit effects: dissipation and dephasing}
\label{Section_Dephasing}

\subsection{Johnson-Nyquist noise of the gate voltage circuit}

\begin{figure}
\epsfysize=12\baselineskip
\centerline{\hbox{\epsffile{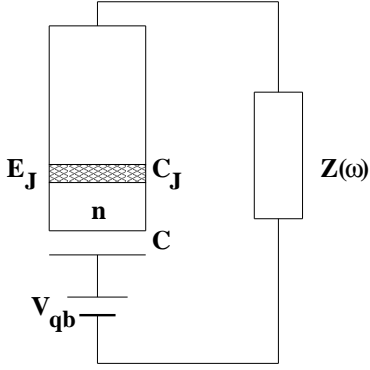}}}
\vskip 0.8cm
\narrowcaption
\caption[]{\label{BITS_CIRCUIT}
Qubit with electromagnetic environment.}
\end{figure}

The idealized picture outlined above has to be extended
to account for the possible dissipation mechanisms
causing decoherence and relaxation.
We focus on the dissipation and fluctuations
which originate  from the circuit of the voltage sources.
In Fig.~\ref{BITS_CIRCUIT} the equivalent circuit of a qubit 
coupled to an impedance $Z(\omega)$ is shown.  The latter is
characterized by intrinsic voltage fluctuations (between its terminals
when disconnected from the circuit), with spectrum   
\begin{eqnarray}
\nonumber
\langle \delta V  \delta V \rangle_\omega & \equiv & 
\int_{-\infty}^\infty dt e^{i\omega t} \frac{1}{2} 
\langle \delta V(t)\delta V(0)+\delta V(0)\delta V(t) \rangle\\
& = & {\rm Re}\{Z(\omega)\} \hbar \omega 
\coth\left( \frac{\hbar \omega}{2k_{\rm B}T}\right) \; .
\label{dVdV_Disconnected}
\end{eqnarray}
When $Z(\omega)$ is embedded in a circuit,
similar to that of Fig.~\ref{BITS_CIRCUIT} but with $E_{\rm J}=0$,
the voltage fluctuations between the terminals of $Z(\omega)$ are 
characterized by a modified spectrum:
\begin{equation}
\langle \delta V  \delta V \rangle_\omega =
{\rm Re}\{Z_{\rm t}(\omega)\} \hbar \omega 
\coth\left( \frac{\hbar \omega}{2k_{\rm B}T}\right) \ .
\label{dVdV_Connected}
\end{equation}
Here $Z_{\rm t}(\omega) \equiv \left[i\omega C_{\rm t} + 
Z^{-1}(\omega)\right]^{-1}$ is the total impedance between 
the terminals of $Z(\omega)$.
Following Caldeira and Leggett~\cite{Caldeira-Leggett} we model 
the dissipative element $Z(\omega)$ by a bath of harmonic oscillators
described by the Hamiltonian
\begin{equation}
        H_{\rm bath} = \sum_\alpha 
        \left[ 
             {1\over {2m_\alpha}}\,p_\alpha^2 
           + {m_\alpha \omega_\alpha^2 \over 2}\,x_\alpha^2
        \right] \ .
\label{Bath_Hamiltonian_xp}
\end{equation}
It is assumed that the voltage between the terminals
of $Z(\omega)$ is given by $\delta V = 
\sum_\alpha \lambda_\alpha x_\alpha$ and the 
spectral function $J(\omega) \equiv {\pi\over 2}
\sum_\alpha {\lambda_\alpha^2 \over 
m_\alpha \omega_\alpha} \delta(\omega - \omega_\alpha)$
is chosen to reproduce the fluctuations spectrum 
(\ref{dVdV_Disconnected}), 
$J(\omega) = \omega {\rm Re}\{Z(\omega)\}$.
Then, we embed the element $Z(\omega)$ in the 
qubit's circuit and we use the Kirchhof constraints
to derive the Hamiltonian of the whole system. 
Taking into account the possibility of a time-dependent 
external voltage $V_{\rm qb}(t)$, we arrive at the resulting 
Hamiltonian   
\begin{eqnarray}
H &=& {(2en)^2\over 2 C_{\rm J}}  
- E_{\rm J} \cos\Theta  
+ \tilde H_{\rm bath} 
\nonumber \\
&-& C_{\rm t} 
\left[ {2en\over C_{\rm J}} + V_{\rm qb}(t)\right]
\sum_\alpha \tilde\lambda_\alpha \tilde x_\alpha 
\ ,
\label{dissHamNewBath}
\end{eqnarray} 
where the tilde-marking of the bath-related quantities 
reflects the fact that the bath has been modified and 
its spectral function corresponds now to the fluctuations 
spectrum (\ref{dVdV_Connected}), 
$\tilde J(\omega) = \omega {\rm Re}\{Z_{\rm t}(\omega)\}$.
Shifting the origins of all the oscillators  
$\tilde x_\alpha \rightarrow \tilde x_\alpha - 
(\tilde \lambda_\alpha/\tilde m_\alpha \tilde \omega_\alpha^2) 
C_{\rm t} V_{\rm qb,0}$, 
where $V_{\rm qb,0}$ is the time 
independent part of the external voltage, we get a more familiar 
form of the Hamiltonian:
\begin{eqnarray}
H &=& {(2en-C_{\rm t}V_{\rm qb,0})^2\over 2 C_{\rm J}}  
- E_{\rm J} \cos\Theta  
+ \tilde H_{\rm bath} 
\nonumber \\
&-& C_{\rm t} 
\left[ {2en\over C_{\rm J}} + \delta V_{\rm qb}(t)\right]
\sum_\alpha \tilde\lambda_\alpha \tilde x_\alpha 
\ ,
\label{dissHamUsualForm}
\end{eqnarray}
To be specific  we will concentrate in the following on the 
fluctuations due to an Ohmic resistor $Z(\omega)=R$ in the 
bias voltage circuit.

\subsection{Relaxation and dephasing rates}

In general, the environment has two effects: inelastic energy relaxation and
dephasing, both characterized by their respective time scales, $\Gamma_{\rm
in}^{-1}$ and $\Gamma_\phi^{-1}$.  To illustrate this point we consider a qubit
prepared in a superposition of eigenstates $a|0\rangle +
b|1\rangle$ with non-vanishing coefficients $a$ and $b$.  This corresponds to 
an
initial density matrix $\hat{\sigma} = \left( ^{|a|^2}_{a^*b} \,
^{ab^*}_{|b|^2}\right)$.  In this case, one question is how fast the diagonal
elements of the density matrix relax to their thermal equilibrium values.  This
relaxation is determined by $\Gamma_{\rm in}$.  The second question is how fast
the off-diagonal elements vanish, which is governed by the rate $\Gamma_\phi$.
In general, the two rates are not equal.

Both inelastic transitions and dephasing were addressed in the context of 
spin-boson 
models, in particular for a two-level system with purely Ohmic dissipation
\cite{Two_Level_Leggett,Two_Level_Weiss}.  Our system (\ref{dissHamUsualForm}) 
reduces to a spin-boson model in the limit where only two charge states need to 
be
considered, with a harmonic oscillator bath coupling to $\sigma_z$. It still
differs from an Ohmic model due to the frequency dependence in the function
$Z_{\rm t}(\omega)$.  However, for the relevant frequencies, $\omega \ll
1/RC_{\rm t}$, the physics is dominated by the resistor. Hence we can take
over the results from the model calculations of
Refs.~\onlinecite{Two_Level_Leggett,Two_Level_Weiss} for purely Ohmic
environments.

The strength of the fluctuations and, hence, the relaxation and 
dephasing effects are controlled by the resistance of the circuit.
It turns out that it has to be measured in units of the  quantum 
resistance $R_{\rm K} \equiv h/e^2 \approx 25.8$k$\Omega$. Hence for low
circuit resistances in the range of $R \approx 50\, \Omega$ we can 
expect a weak effect. Furthermore, in
our system the effect of fluctuation is reduced
due to the weak capacitive coupling of the qubit to the circuit. 
Indeed, as is apparent from (\ref{dissHamUsualForm}) the coupling of 
$\sum_\alpha \tilde \lambda_\alpha \tilde x_\alpha$ to the qubit's charge 
$n$  involves the ratio $C/(C_{\rm J} +C)= C_{\rm t}/ C_{\rm J} \ll 1$. 
Thus, the relevant parameter characterizing the effect of the voltage
fluctuations is 
$(R/R_{\rm K}) \left( C_{\rm t}/C_{\rm J}\right)^2$.

In Ref.~\onlinecite{Two_Level_Leggett} only the unbiased case of the 
spin-boson model, corresponding to the qubit at the degeneracy point 
$n_{\rm qb} = 1/2$, has been analyzed rigorously in the limit of low 
dissipation (the only limit relevant for the quantum computation).
The biased case of the spin-boson model at low dissipation 
(i.e.\ the qubit out of the degeneracy)
was treated in Ref.~\onlinecite{Two_Level_Weiss}. 
The two rates, expressed in terms of the mixing 
angle introduced in (\ref{Eigen_Basis}), are
\begin{equation}
\label{Gamma_r_Weiss}
\Gamma_{\rm in} = 
\sin^2\eta \, 4\pi{R \over R_{\rm K}}
 \left({C_{\rm t} \over  C_{\rm J}}\right)^2 {\Delta E\over \hbar}
\coth{\Delta E\over 2k_{\rm B} T} \ ,
\end{equation}
and
\begin{equation}
\Gamma_{\phi} =
{1 \over 2} \, \Gamma_{\rm in} +  \cos^2\eta \; 8\pi {R \over R_{\rm K}}
\left({C_{\rm t}\over C_{\rm J}}\right)^2 {k_{\rm B} T\over \hbar} \ .
\label{Gamma_phi_Weiss}
\end{equation}
We note that out of the degeneracy the dephasing rate acquires a
component proportional to the temperature. It should also be noted 
that the relaxation rate (\ref{Gamma_r_Weiss}) may be obtained in 
a much simpler way~\cite{Our_PRL} using the so called ``$P(E)$'' 
theory discussed in 
Refs.~\cite{P(E)_Panyukov_Zaikin,P(E)_Odintsov,P(E)_Nazarov,P(E)_Devoret}.  

Thus, the diagonal and off-diagonal elements
in the basis of the qubit's eigenstates relax 
with different rates described above, while the 
off-diagonal elements carry also an oscillating phase 
factor. The last is translated in the charge basis to the
`coherent charge oscillations', which are observed in the 
quantity $\langle\sigma_z(t)\rangle$. In the absence of dissipation 
this quantity oscillates coherently around some average value. With 
dissipation, this average value relaxes to equilibrium with rate 
$\Gamma_{\rm in}$, while the oscillations decay with rate 
$\Gamma_{\phi}$~\cite{Two_Level_Weiss}.

The factors $\sin^2\eta$ and $\cos^2\eta$ in 
Eqs.~(\ref{Gamma_r_Weiss},\ref{Gamma_phi_Weiss}) 
indicate  the nature of different
contributions to  the relaxation and dephasing rates. 
In the basis of the eigenstates (\ref{Eigen_Basis}) the qubit's part of the
Hamiltonian (\ref{dissHamUsualForm}) is diagonal. However, 
the circuit electromagnetic fluctuations couple to the charge.
Hence, in the notation introduced after (\ref{Eigen_Basis})
we have
\begin{eqnarray}
\label{Fluctuation_Hamiltonian}
H_{\rm em} &=& - {C_{\rm t} \over C_{\rm J}} 2e\, \sigma_z 
\sum_\alpha {\tilde \lambda_\alpha \tilde x_\alpha}
\nonumber \\ 
&=& 
- {C_{\rm t} \over C_{\rm J}} 2e\, (\cos\eta\,\rho_z+
\sin\eta\,\rho_x) \sum_\alpha {\tilde \lambda_\alpha \tilde x_\alpha}
\ .
\end{eqnarray}  
Since the first term commutes with the qubit's Hamiltonian,
only the second one in (\ref{Fluctuation_Hamiltonian})
contributes to the transitions between the eigenstates.
This explains the factor $\sin^2\eta$ in (\ref{Gamma_r_Weiss}).
The first term in (\ref{Fluctuation_Hamiltonian}),
while ineffective for transitions, makes the energy difference
between the qubit's levels fluctuate. Therefore, the 
off-diagonal elements of the density matrix acquire a random 
phase. The last process is ``pure'' dephasing.
It occurs even when transitions are suppressed (e.g.\ when
$E_{\rm J} = 0$).  This particular case 
($E_{\rm J} = 0$, $\delta V_{\rm qb}(t)=0$) is quite simple 
and we can reproduce Eq.~(\ref{Gamma_phi_Weiss}) in an exact
calculation, examining also the zero temperature limit. 
To do this, we note that in the two state approximation the 
off-diagonal element of the reduced density matrix is given by 
$\sigma_{0,1}(t)=\langle\exp(-i\Theta(t))\rangle$. 
To calculate it we perform a canonical transformation
$\tilde x_\alpha' = \tilde x_\alpha - 
(C_{\rm t}/C_{\rm J})
({\tilde\lambda_\alpha / \tilde m_\alpha \tilde \omega_\alpha^2})
\,2en$,
$\Theta' = \Theta + (2\pi/\Phi_0)(C_{\rm t}/C_{\rm J})\,
\sum_\alpha ({\tilde\lambda_\alpha / \tilde m_\alpha \tilde \omega_\alpha^2})
\tilde p_\alpha$,
after which the Hamiltonian (\ref{dissHamUsualForm}) becomes: 
\begin{eqnarray}
H &=& {(2en-C V_{\rm qb,0})^2\over 2 (C+C_{\rm J})}
\nonumber \\  
        &-& E_{\rm J} \cos\left(\Theta' - 
        {2\pi\over \Phi_0}{C_{\rm t}\over C_{\rm J}} 
        \sum_\alpha 
        {\tilde\lambda_\alpha \over \tilde m_\alpha \tilde \omega_\alpha^2}
        \tilde p_\alpha\right)
\nonumber \\
&+& H_{\rm bath}
- C_{\rm t} \delta V_{\rm qb}(t) 
\left[\sum_\alpha {\tilde \lambda_\alpha \tilde x_\alpha'}
      +{2en \over C_{\rm J}}
\right] 
\ .
\label{dissHamPureDephasing}
\end{eqnarray}
The Josephson and the last terms are zero in our case and 
the evolution of \{$\Theta',n$\} is decoupled from the bath.
In the two state approximation ($n^2 = n$) the equations 
of motion read: $d\,n/dt = 0$, 
$d\,[\exp(-i\Theta')]/dt = -i E_{\rm ch} (V_{\rm qb,0}) \exp(-i\Theta')$,
where $E_{\rm ch}(..)$ was introduced in Eq.~(\ref{Magnetic_Hamiltonian}).
Thus, introducing 
$\Phi \equiv (2\pi/\Phi_0)(C_{\rm t}/C_{\rm J})
\sum_\alpha ({\tilde\lambda_\alpha / \tilde m_\alpha \tilde \omega_\alpha^2})
\tilde p_\alpha$,
we get
\begin{eqnarray}
\label{ZeroEJ_Dephasing}
&&\sigma_{0,1}(t) = \langle \exp(i\left[\Phi(t)-\Theta'(t)\right]) \rangle 
\nonumber \\
&&= \langle \exp(i\Phi(t))\,
      \exp(-i\left[\Theta'(0)+E_{\rm ch}(V_{\rm qb,0})\,t\right]) \rangle      
\nonumber \\
&&= \sigma_{0,1}(0)\,e^{-iE_{\rm ch}(V_{\rm qb,0})\,t}\,
     \langle e^{i\Phi(t)}\,e^{-i\Phi(0)} \rangle 
\ .
\end{eqnarray} 
The correlator $\langle e^{i\Phi(t)}\,e^{-i\Phi(0)} \rangle$ was studied 
extensively by many authors 
\cite{P(E)_Panyukov_Zaikin,P(E)_Odintsov,P(E)_Nazarov,P(E)_Devoret}. 
It is equal to $\exp(4(C_{\rm t}/C_{\rm J})^2 J(t))$, where
\begin{eqnarray}
\label{J(t)}
	&&J(t) =  
	2\int_0^{\infty} {d\omega \over\omega} 
	{Re Z_{\rm t}(\omega)\over R_K}  \nonumber \\
	&&\times
	\left[\coth\left({\hbar\omega\over2 k_B T}\right)[\cos(\omega t)-1]
	-i\sin(\omega t)\right] \ . 
\end{eqnarray}
Since ${\rm Re}Z_{\rm t}(\omega) = R/(1+R^2C_{\rm t}^2\omega^2)$, 
we may roughly substitute the bath's spectrum by a purely Ohmic one with a 
cut-off set at $\omega_{\rm c} \equiv \omega_{RC} = 1/(RC_{\rm t})$. Then, 
at non-zero temperature, ${\rm Re}J(t) 
\approx -(2\pi k_{\rm B}T/\hbar)(R/R_{\rm K})\,t$
for $t>\hbar/2k_{\rm B}T$, and we reproduce Eq.~(\ref{Gamma_phi_Weiss})
in the limit $E_{\rm J} = 0$. At zero temperature 
${\rm Re} J(t) \approx  -(2R/R_{\rm K})\,\log(\omega_{\rm c}t)$
for $t > 1/\omega_{\rm c}$. 
Thus, the fact that the ``pure'' dephasing rate vanishes at zero 
temperature does not mean that there is no dephasing at all. 
The decay of the off-diagonal element of the reduced density matrix 
is just non-exponential but rather algebraic
(see Refs.~\onlinecite{Hakim_Ambegaokar,Golubev_Zaikin_Dephasing}).  

We conclude that the dephasing of the initial quantum 
state of the qubit is caused by two different processes:
the dissipative transitions between the eigenlevels,
and the fluctuations of the energy difference between the levels.
The off-diagonal elements of the density matrix are suppressed
by both of them. While the first process survives at low temperature,
the second one produces the dephasing rate proportional to the 
temperature.

These results should be taken into account when selecting the optimum 
idle state for the qubit. If the temperature is low,
$k_{\rm B} T \ll E_{\rm J}$, 
the best choice is obviously far from the degeneracy 
point ($\sin \eta \ll 1$). 
Also at higher temperatures this regime has a longer coherence time
than the degeneracy point, although only by a numerical factor.
In all cases it is pure dephasing, rather than inelastic transitions
which limits the coherence at the idle point.


\subsection{Discussion and extensions}

In summary, the dephasing rate is small if the effective resistance of the
circuit is low compared to the quantum resistance, $R_{\rm K}$.  Furthermore, a
low gate capacitance $C$ reduces the coupling of the qubit to the environment.
Hence, with suitable parameters ($R \le 50\Omega, C/C_{\rm J} \le 0.1$) at low
temperatures the number of operations which can be performed before the
environment destroys the phase coherence may be as large as $10^3$--$10^4$.  
The
value of the phase coherence time in this case is of the order of 10--100 ns.
Coherence times of this order of magnitude have been observed in experiments on
quantum dots~\cite{Kouwenhoven}.

Longer phase coherence times would be achieved if the coupling of the qubit to
the noisy environment could be further reduced, without weakening the coupling
to the common inductor $L$ which provides the 2-bit coupling.  The model
discussed above is not ideal since the Ohmic resistor with Johnson-Nyquist 
noise
is assumed to be in the immediate vicinity of the qubit, reduced in its effect
only by a low gate capacitor.  A suitable design with a combination of
superconducting leads and filters can drastically improve the situation.

If the physical dephasing effects are reduced, the more serious will be the
errors related to imperfections in the time control or imprecise parameters and
coupling energies, e.g., non-vanishing 2-bit couplings during the idle 
periods.  
If the combination of all these effects is sufficiently reduced, allowing for
$10^4$--$10^5$ coherent manipulations steps, then eventually the remaining
errors can be corrected by suitable codes~\cite{errorcorrectingcodes}. It 
appears from our analysis that this goal can be reached with Josephson 
junction qubits.


\section{Measuring the state of the qubit}
\label{Section_Measurement}

\subsection{The model for a  single-electron transistor attached to the qubit}

\begin{figure}  
\epsfysize=18\baselineskip
\centerline{\hbox{\epsffile{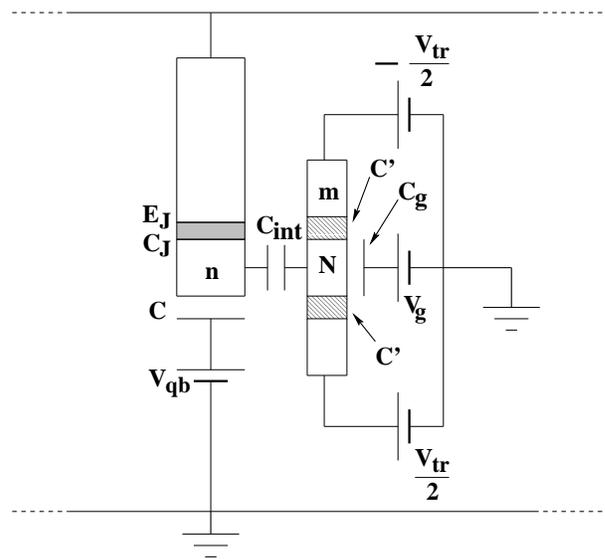}}}
\vspace{5mm}
\narrowcaption
\caption[]{\label{CIRCUIT}
The circuit consisting of a qubit plus a SET transistor
used as a measuring device.}
\end{figure}

The read-out of the state of the qubit requires a
quantum measurement process. Since the relevant quantum degree 
of freedom is the charge of the qubit island, the natural choice of 
measurement device is a single-electron transistor (SET). 
This system is shown in Fig.~\ref{CIRCUIT}.
The left part is the qubit, with state  characterized by
the number of extra Cooper pairs, $n$, on the island,
and controlled by its gate voltage, $V_{\rm qb}$. 
The right part shows a normal island between two normal leads, 
which form the SET. It's charging state is characterized by the 
number of extra single-electron charges on the middle island, $N$. 
It is controlled by gate and transport voltages, $V_{\rm g}$
and $V_{\rm tr}$, and further, due to
the capacitive coupling to the qubit, by the state of the latter.
A similar setup has been studied in the
experiments of Refs.~\onlinecite{Bouchiat_PhD}
with the purpose to demonstrate that the ground
state of a single Cooper pair box is a coherent 
superposition of different charge states. We will discuss the 
relation of  these experiments to our proposal below.

During the quantum manipulations of the qubit the transport voltage $V_{\rm 
tr}$
across the SET transistor is kept zero and the gate voltage of the SET is 
chosen
to tune the island away from degeneracy points.  Therefore no dissipative
currents flow in the system, and the transistor merely modifies the 
capacitances
of the system.  To perform a measurement one tunes the SET by $V_{\rm g}$ to 
the
vicinity of its degeneracy point and applies a small transport voltage $V_{\rm
tr}$.  The resulting normal current through the transistor depends on the 
charge
configuration of the qubit, since different charge states induce different
voltages on the middle island of the SET transistor.  In order to investigate
whether the dissipative current through the SET transistor allows us to resolve
different quantum states of the qubit, we have to discuss various noise 
factors,
including the shot noise associated with the tunneling current and the
measurement induced transitions between the states of the qubit.  For this
purpose we analyze the time evolution of the density matrix of the combined
system.  We find that for suitable parameters, which can be realized
experimentally, the dephasing by the passive SET is weak.  When the transport
voltage is turned on the dephasing is fast, and the current through the
transistor --- after a transient period --- provides a measure of the state of
the qubit.  At still longer times the dynamics of the SET destroys the
information of the quantum state to be measured.

The Hamiltonian of the combined system 
\begin{eqnarray}
\label{totalH}
	H = & H_{\rm ch} + H_{\rm L} + H_{\rm R} + H_{\rm I}
\nonumber\\ 
	 & + H_{\rm J} + H_{\rm T}  
\end{eqnarray}
contains the charging energy, the terms describing the 
microscopic degrees of freedom of the metal islands and electrodes, 
and the tunneling terms, including the Josephson
coupling. The charging term is a quadratic form in the 
variables $n$ and $N$. I.e. for the system shown in Fig.~\ref{CIRCUIT} it is
\begin{eqnarray}
\label{CHARGING_ENERGY}
H_{\rm ch} (n,N,V_n,V_N)&=&4E_{\rm qb} n^2 + E_{\rm SET} N^2 + 
2E_{\rm int} nN 
\nonumber \\
&& + 2en V_{n} + eN V_{N} \, .
\end{eqnarray} 
The charging energy scales $E_{\rm qb}$, $E_{\rm SET}$ and $E_{\rm int}$ are
determined by the capacitances between all the islands. 
Introducing 
$$
A \equiv (C+C_{\rm J})(C_{\rm g} + C_{\rm int} + 2C') + 
C_{\rm int}(C_{\rm g}+2C') \approx 2 C_{\rm J} C' \ ,
$$
we can write them as
\begin{eqnarray}
\label{CHARGING_ENERGY2}
E_{\rm qb} &=& e^2(C_{\rm g} + C_{\rm int} + 2C')/ 2A \approx e^2/2C_{\rm J}
\ ,\nonumber\\
E_{\rm SET} &=& e^2(C+C_{\rm int}+C_{\rm J})/2A \approx  e^2/(4C')
\ ,\nonumber\\
E_{\rm int} &=& e^2C_{\rm int}/ A \approx e^2C_{\rm int}/(2C_{\rm J} C')
\ .
\end{eqnarray} 
Here we have assumed that the two junctions of the SET have equal
capacitances $C'$, and the approximate results refer to the limit
$C,C_{\rm int},C_{\rm g} \ll C' \ll C_{\rm J}$, which we consider useful (see 
below).
The effective gate voltages  $V_{n}$ and $V_{N}$
depend in general on all  voltages $V_{\rm qb}$, $V_{\rm g}$, and 
the voltages applied to both electrodes of the SET. However,
for a symmetric setup (equal junction capacitances)
and symmetric distribution of the transport voltage, $V_{\rm tr}$,  
between both electrodes of the SET (as shown in Fig.~\ref{CIRCUIT}),  
$V_{n}$ and $V_{N}$ are controlled only by the two gate voltages,
\begin{eqnarray}
\label{Effective_Voltages}
&&V_N = V_{\rm g} {C_{\rm g}(C+C_{\rm int}+C_{\rm J})\over A} +
              V_{\rm qb} {C_{\rm int}C\over A}
\ ,\nonumber \\
&&V_n = V_{\rm g} {C_{\rm g}C_{\rm int}\over A} +
              V_{\rm qb}{(C_{\rm g} +C_{\rm int} + 2 C')C\over A}
\ .
\end{eqnarray}

The microscopic terms describe 
noninteracting electrons in the two leads and on the middle island 
of the SET transistor
\begin{equation}
\label{FREE_LEADS_H}
        H_r =  \sum_{k\sigma} \epsilon_{k\sigma}^{r} c_{k\sigma}^{r\dag} 
        c_{k\sigma}^{r \phantom{\dag}}
        \ \ (r= {\rm L,R,I})\ . 
\end{equation}
The index $\sigma$ labels transverse channels including the spin,
while $k$ labels the wave vector within one channel. 

Similar terms describe the electrode and island of the qubit; however, for
the superconducting non-dissipative element  the microscopic degrees of
freedom can be integrated out~\cite{Schon_Zaikin_Review}, resulting in the 
``macroscopic'' quantum description presented in Sect.~II.
In this limit the tunneling terms reduce to the 
Josephson coupling $H_{\rm J} = - E_{\rm J}\cos\Theta$, expressed
in a collective variable describing the coherent transfer 
of Cooper pairs in the qubit, $e^{i\Theta} |n\rangle = |n+1\rangle$.
 
The normal-electron tunneling in the SET transistor is described by the
standard tunneling Hamiltonian, which 
couples the microscopic degrees of freedom, 
\begin{eqnarray}
\label{TUNNEL_HAMILTONIAN}
        H_{\rm T} = & \sum\limits_{kk'\sigma} 
        T^{L}_{kk'\sigma} c^{\rm L\dag}_{k\sigma}
        c_{k'\sigma}^{\rm I \phantom{\dag}} e^{-i\phi} 
\nonumber \\ 
        &  + \sum\limits_{k'k''\sigma} 
        T^{R}_{k'k''\sigma} c^{\rm R\dag}_{k''\sigma} 
        c_{k'\sigma}^{\rm I \phantom{\dag}} e^{-i\phi} e^{i\psi}
        + {\rm h.c.} \ .  
\end{eqnarray} 
To make the charge transfer explicit, (\ref{TUNNEL_HAMILTONIAN})
displays two ``macroscopic''  operators, $e^{\pm i\phi}$ and $e^{\pm i\psi}$.
The first one describes changes of the charge on the transistor island 
due to the tunneling: $e^{i\phi} |N\rangle = |N+1\rangle$. It may be treated 
as an independent degree of freedom if the total
number of electrons on the island is large. We further include the
operator $e^{\pm i\psi}$ which describes the changes of the charge
in the right lead. It acts on $m$, the 
number of electrons which have tunneled through the SET transistor, 
$e^{i\psi} |m\rangle = |m+1\rangle$. 
Since the chemical potential
of the right lead is controlled, $m$ does not appear in any charging 
part of the Hamiltonian.
However, we have 
to keep track of it, since it is the measured quantity,
related to the current through the SET transistor.

In equilibrium, i.e. when the SET is kept in the state  $N=0$,
the qubit's dynamics is described by the same  Hamiltonian 
as discussed in previous sections,
$H_{\rm qb} = 4E_{\rm qb} (n - n_{\rm qb})^2 - E_{\rm J} \cos \Theta$,
where $n_{\rm qb} \equiv -eV_n/4E_{\rm qb}$.
We recall that in the limit where  only the lowest energy charge  
states $n=0$ and $n=1$ are relevant, it 
reduces to a two state quantum system. In the basis of eigenstates
(\ref{Eigen_Basis}), $|0\rangle$ and $|1\rangle$, 
which are expressed in terms of the mixing angle $\eta$, where
$\tan\eta = E_{\rm J} / E_{\rm ch}(n_{\rm qb})$, 
it becomes $H_{\rm qb} = \frac{1}{2}\Delta E_\eta\,\rho_{z}$, 
where $\Delta E_\eta \equiv E_{\rm J}/\sin\eta$ and
$E_{\rm ch}(n_{\rm qb}) \equiv 4E_{\rm qb}(1-2n_{\rm qb})$.
In this basis the number operator $n$ becomes non-diagonal,
\begin{equation}
\label{n_OPERATOR}	       
	n={1\over 2}(1+\sigma_z)= 
          {1\over 2}(1+\cos\eta\,\rho_z+\sin\eta\,\rho_x) \ .
\end{equation} 
For the following discussion we choose $n_{\rm qb}$ away from the
degeneracy point, which combined with  $E_{\rm J} \ll E_{\rm qb}$ 
implies $\tan\eta \ll 1$.

The  mixed term in (\ref{CHARGING_ENERGY}) provides the
 interaction Hamiltonian. With $n$ given by (\ref{n_OPERATOR}) it becomes 
\begin{equation}
\label{INTERACTION_HAMILTONIAN}
        H_{\rm int} = E_{\rm int} N\,\sigma_z 
                    = E_{\rm int} N (\cos\eta\,\rho_z + \sin\eta\,\rho_x) \ ,
\end{equation} 
plus an extra term, $E_{\rm int} N$, which together with
further terms is collected in the Hamiltonian of the SET 
transistor,
\begin{equation}
\label{SET_HAMILTONIAN}
        H_{\rm SET} = E_{\rm SET} (N - N_{\rm SET})^2  
        + H_{\rm L} + H_{\rm R} + H_{\rm I} + H_{\rm T} \ .
\end{equation}
The transistor's gate charge  became
$N_{\rm SET} \equiv -(eV_{N} + E_{\rm int}) / 2E_{\rm SET}$.
The total Hamiltonian thus reads
\begin{equation}
H = H_{\rm qb} + H_{\rm SET} + H_{\rm int} \ .
\end{equation} 

The total system composed of qubit and SET is described by a total 
density matrix $\hat{\rho}(t)$, which we can reduce, by taking a trace 
over the microscopic states of the left and right
leads and of the island, to 
$\hat{\sigma}(t) = {\rm Tr}_{\rm L,R,I} \{\hat{\rho}(t)\}$.
In general, this reduced density matrix  $\hat{\sigma}(i,i';N,N';m,m')$ 
is a matrix in the index $i$, which stand
for the quantum states of the qubit 
$|0\rangle$ or $|1\rangle$, in $N$, and in $m$.
In the following we will assume that initially --- as a result of previous 
quantum manipulations --- the qubit is prepared in the quantum
state \,$a|0\rangle + b|1\rangle$, and at time $t=0$ we switch on 
a transport voltage to the single-electron transistor.
We then proceed to further reduce the density 
matrix in two ways which provide complementary information 
about the measuring process. 

The first, widely used procedure \cite{Gurvitz} is to trace
over $N$ and $m$. This yields a reduced density matrix of
the qubit $\sigma_{i,j} \equiv \sum_{N,m} \hat{\sigma}(i,j;N,N;m,m)$.
Just before the measurement, it is in the state
\begin{equation}
\label{SIGMA(0)}
\hat{\sigma}(0) = 
\left(
\begin{array}{cc}
|a|^2 & a b^* \\
a^* b & |b|^2
\end{array}
\right) \ .
\end{equation}
The questions then arise  how fast the off-diagonal elements of
$\hat{\sigma}_{i,j}$ vanish, i.e.\ how fast is the dephasing, and how fast
the diagonal elements change their values after the SET is switched to the
dissipative state.
These questions are the same as in our discussion of fluctuation effects 
in Section \ref{Section_Dephasing}. However, the coupling to the 
dissipative SET in general shortens these times. This description 
is enough when one is interested in the quantum properties
of the measured system, i.e.\ the qubit only, and the 
measuring device serves merely as a source of dephasing
\cite{Levinson,Aleiner,Gurvitz,Buks}. It does not tell us much
about the quantity measured in an experiment,
namely the current flowing through the SET. 

The second procedure, pursued in the following,
is to evaluate the probability distribution of
the number of electrons $m$ which tunnel through the SET
during time $t$,
\begin{equation}
\label{P(m)}
        P(m,t) \equiv \sum_{N,i} \hat{\sigma}(i,i;N,N;m,m)(t) \ .  
\end{equation}
This distribution provides the information about the experimentally  
accessible quantity during the measurement process.
At $t=0$ no electrons have tunneled, so $P(m,0) = \delta_{m,0}$.
Then the peak of the distribution moves in positive $m$-direction 
and, simultaneously, it widens due to shot noise.
Since two states of the qubit correspond to different 
tunneling currents, and hence shift velocities in $m$-direction, 
one may hope that after some time the 
peak splits into two. If after sufficient separation of the two
peaks their weights are still close to $|a|^2$
and $|b|^2$, a good quantum measurement can be performed by measuring
$m$. After a longer time further processes destroy this idealized picture.
The two peaks transform into a broad plateau, since transitions between the 
qubit's states are induced by the measurement. 
Therefore, one should find an optimum
time for the measurement, such that, on one hand, the two peaks are separate
and, on the other hand, the induced transitions have not yet influenced
the process.
In order to describe this  we have to 
analyze the time evolution of the reduced density matrix quantitatively.

\subsection{Quantitative description of the measurement process}

\begin{figure}  
\epsfysize=3.5\baselineskip
\centerline{\hbox{\epsffile{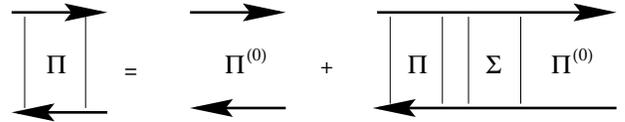}}}
\vskip 0.8cm
\narrowcaption
\caption[]{\label{DYSON}
The Dyson-type equation governing the time evolution of the density matrix.
It is equivalent to the generalized master equation (\ref{MASTER_EQUATION}).
The `self energy' diagrams $\Sigma$ describes the transitions due
to tunneling in the SET transistor.}
\end{figure}

The time evolution of the density matrix leads to 
Bloch-type master equations with coherent terms. Examples
have recently been analyzed in contexts similar to the present
\cite{Schoeller_PRB,Nazarov,Gurvitz}. 
In Ref.~\onlinecite{Schoeller_PRB} a diagrammatic
technique has been developed which provides a formally exact
master equation as an expansion in the tunneling 
term $H_{\rm T}$,
while all other terms constitute the zeroth order Hamiltonian
$H_0 \equiv H - H_{\rm T}$, which is treated exactly.
The time evolution of the reduced density matrix is given by 
$\hat\sigma(t) = \hat\sigma(0)\Pi(0,t)$. 
The propagator $\Pi(t',t)$ can be expressed in a 
diagrammatic form and finally summed up in a way reminiscent 
of a Dyson equation. Examples are shown in Fig.~\ref{DYSON}. 
In contrasts to ordinary many-body expansions, since the time dependence
of the density matrix is described by a forward and a backward 
time-evolution operator, there are two propagators, which are 
represented by  two horizontal lines (Keldysh contour). The two
bare lines describe the coherent time evolution of the system. They
are coupled due to the tunneling in the SET, which is treated as a
perturbation. The sum of 
all distinct transitions defines a `self-energy' diagram $\Sigma$.
Below we will present the rules how to calculate  $\Sigma$ and 
present a suitable approximate form. The Dyson equation is equivalent to
a (generalized) master equation for the density matrix, which reads
\begin{equation}
\label{MASTER_EQUATION}
        {d\hat{\sigma}(t)\over dt} - 
        {i\over\hbar}[\hat{\sigma}(t), H_0]
        = \int_0^t dt' \hat{\sigma}(t') \Sigma(t-t') \ .
\end{equation}
In the present problem
the density matrix is a matrix $ \hat{\sigma}(i,i';N,N';m,m') \equiv
\hat{\sigma}^{i', N',m'}_{i, \; N,\; m}$ 
in all three indices $i$, $N$, and $m$, and the  (generalized)
transition rates due to single-electron tunneling processes
(in general of arbitrary order),
$\Sigma^{i',N',m' \rightarrow \bar{i'}, \bar{N'},\bar{m'}}
_{i\phantom{'},N\phantom{'},m\phantom{'}  \rightarrow \,  \bar{i}, \; 
\bar{N},\; 
\bar{m}}(t - t')$,
connect these diagonal and off-diagonal states. 

\begin{figure}  
\epsfysize=5\baselineskip
\centerline{\hbox{\epsffile{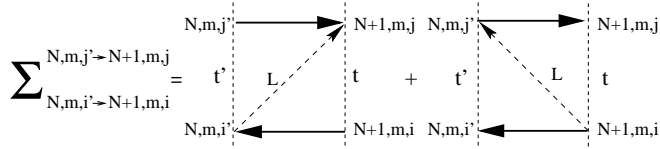}}}
\narrowcaption
\vskip1mm
\caption[]{\label{DIAGRAM_GAIN}
Example of a  `self energy' diagram for an ``in'' rate.}
\end{figure}

\begin{figure}
\epsfysize=5\baselineskip
\centerline{\hbox{\epsffile{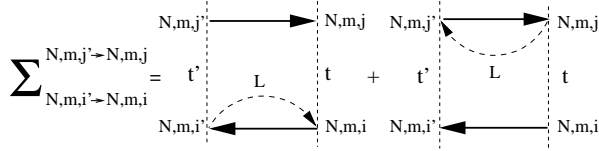}}}
\narrowcaption
\vskip1mm
\caption[]{\label{DIAGRAM_LOOSE}
Example of a  `self energy' diagram for an ``out'' rate.}
\end{figure}

The transition rates can be calculated diagrammatically in the framework of the
real-time Keldysh contour technique.  We briefly review the rules for their
evaluation; for more details including the discussion of higher order diagrams
we refer to Ref.~\onlinecite{Schoeller_PRB}.  Typical diagrams, which will be
analyzed below, are displayed in Fig.~\ref{DIAGRAM_GAIN}
and~\ref{DIAGRAM_LOOSE}.  Again the horizontal lines describe the time 
evolution
of the system governed by the zeroth order Hamiltonian $H_0$.  Their properties
will be discussed below.  The directed dashed lines stand for tunneling
processes, in the example considered the tunneling takes place in the left
junction.  According to the rules the dashed lines
contribute the following factor to the self-energy $\Sigma$,
\begin{equation}
\label{LINE}
        \alpha_{\rm L}\, \left({\pi k_{\rm B}T\over\hbar}\right)^2 \, 
	{\exp\left[\pm \frac{i}{\hbar}\mu_{\rm L}(t-t')\right]
        \over
        \sinh^2\left[{\pi k_{\rm B}T\over\hbar}
               \left(t-t' \pm i\delta\right)\right]} 
\ , 
\end{equation}
where $\alpha_{\rm L} \equiv \hbar / (4\pi^2 e^2 R_{\rm T,L})$
is the dimensionless tunneling conductance,
$\mu_{\rm L}$ is the electro-chemical potential of the left lead, and
$\delta^{-1}$ is the high-frequency cut-off, which is at most of order
of the Fermi energy. The sign of the infinitesimal term $i\delta$  
depends on the time-direction of the dashed line.
It is negative if the direction of the line with respect 
to the Keldysh contour coincides with its direction 
with respect to the absolute time (from left to right), and positive
otherwise. For example the right 
part of Fig.~\ref{DIAGRAM_GAIN} should carry a minus sign, while the left part 
carries a plus sign. Furthermore, the sign in front of  $i\mu_{\rm L}(t-t')$ 
is negative (positive), if the line goes forward (backward)
with respect to the absolute time. The first order diagrams are multiplied
by $(-1)$ if the dashed line connects two points on different branches of 
the Keldysh contour.   

The horizontal lines describe the time-evolution of the system between
tunneling processes. For an isolated transistor island they reduce
to simple exponential factors $e^{\pm \frac{i}{\hbar}E(t-t')}$, depending on 
the charging 
energy of the system. 
In the present case, however, where the island is coupled to the qubit
we have to account for the nontrivial time evolution of the latter.
For instance,  the upper line in the left part
of Fig.~\ref{DIAGRAM_GAIN} corresponds to 
$\langle N,j|e^{-\frac{i}{\hbar}H_0(t-t')}|N,j'\rangle$,
while the lower line corresponds to 
$\langle N+1,i'|e^{\frac{i}{\hbar}H_0(t-t')}|N+1,i\rangle$.

In principle the density matrix is an arbitrary non-diagonal matrix in all 
three
indices $i$, $N$, and $m$.  But, as has been shown in
Ref.~\onlinecite{Schoeller_PRB}, a closed set of equations can be derived,
describing the time evolution of the system, which involves only the diagonal
elements in $N$.  The same is true for the matrix structure in $m$.  Therefore,
we need to consider only the following elements of the density matrix
$\hat{\sigma}^{i,N,m}_{j,N,m}$.  Accordingly, of all the transition rates we
need to calculate only those between the corresponding elements of the density
matrix, i.e.\ $\Sigma^{i',N',m' \rightarrow i, N,m}_{j',N',m' \rightarrow
j,N,m}(\Delta t)$.  In the present problem we further can assume that the
tunneling conductance of the SET is low compared to the inverse quantum
resistance.  In this case, lowest order perturbation theory in the single
electron tunneling, describing `sequential tunneling processes', is sufficient.
The diagrams for $\Sigma$ can be split into two classes, depending on whether
they provide expressions for off-diagonal ($N'\ne N$) or diagonal ($N'=N$) in
$N$ elements of $\Sigma$.  In analogy to the scattering integrals in the
Boltzmann equation these can be labeled ``in'' and ``out'' terms, in the sense
that they describe the increase or decrease of a given element
$\hat{\sigma}^{i,N,m}_{j,N,m}$ of the density matrix due to transitions from or
to other $N$-states.  Examples for the ``in'' and ``out'' terms are shown in
Fig.~\ref{DIAGRAM_GAIN} and Fig.~\ref{DIAGRAM_LOOSE}, respectively.

We now are ready to evaluate the rates in Fig.~\ref{DIAGRAM_GAIN} and
Fig.~\ref{DIAGRAM_LOOSE}.  As an example we consider an ``in'' tunneling 
process 
in the left junction:
\begin{eqnarray}
\label{LEFT_RATE_MATRIX}
        &&\Sigma^{j',N,m \rightarrow j,N+1,m}
	_{i',N,m \rightarrow i,N+1,m}(\Delta t)= 
        -\alpha_{\rm L}\left({\pi k_{\rm B}T\over\hbar}\right)^2
        \nonumber \\
        &&\times \left\{
        {\exp\left[-\frac{i}{\hbar}
        	(\tilde E^{N}_{N+1}+W^{N}_{N+1}) \Delta t
             \right]  
        \over
        	\sinh^2\left[
        		{\pi k_{\rm B}T\over\hbar}(\Delta t +
        			i\delta)
        	\right]
        }
        \right.
        \\
        &&
        \left.
        +
        {	\exp\left[
        		-\frac{i}{\hbar} (\tilde E^{N+1}_{N}+W^{N+1}_{N})
        		\Delta t
        	\right]  
        \over
        	\sinh^2\left[
        		{\pi k_{\rm B}T\over\hbar}(\Delta t -
        			i\delta)
        	\right]
        }
        \right\}^{j'j}_{i'i} 
        \nonumber
        \ ,
\end{eqnarray}  
where $\tilde E^{N_1}_{N_2} \equiv
(E_{\rm SET}(N_1)- \mu_{\rm L}\,N_1) - (E_{\rm SET}(N_2)-\mu_{\rm L}\,N_2)$
is the usual Coulomb
energy gain for the tunneling in the left junction in the absence of the qubit, 
and
\begin{equation}
W^{N_1}_{N_2} =
H_{\rm qb}^{\rm T}(N_1)\otimes 1 - 1\otimes H_{\rm qb}(N_2)\ .
\label{W_N1_N2}
\end{equation}
provide corrections to the energy gain sensitive to the state of the qubit.
Here $H_{\rm qb}(N)$ is the $N$-th block of the Hamiltonian 
$H_{\rm qb}+H_{\rm int}$ (note that $H_{\rm qb}$ and $H_{\rm int}$
are block-diagonal with respect to $N$). The indices $j',j$ and $i',i$ relate 
to the left and right side of the tensor product in (\ref{W_N1_N2})
correspondingly. 

The form of the master equation (\ref{MASTER_EQUATION}) suggests 
the use of the Laplace transform, after which the last
term in (\ref{MASTER_EQUATION}) becomes $\Sigma(s)\hat{\sigma}(s)$. 
We Laplace transform (\ref{LEFT_RATE_MATRIX}) in the regime 
$\hbar s, |W^{N}_{N+1}|,|W^{N+1}_{N}| \ll \tilde E^{N}_{N+1}$, i.e.\ 
we assume the density matrix $\hat{\sigma}$ to change slowly on the time 
scale given by $\hbar/\tilde E^{N}_{N+1}$. This assumption should be 
verified later for self-consistency. These inequalities also mean that
we choose the operation regime of the SET far enough from the Coulomb 
threshold. Therefore, the tunneling is either ``allowed'' for both states
of the qubit or it is ``blocked'' for both of them. At low temperatures 
($k_{\rm B}T \ll \tilde E^{N}_{N+1}$) and for 
$\tilde E^{N}_{N+1}\delta \ll \hbar$ we obtain:
\begin{eqnarray}
\label{LEFT_RATE_LAPLACE}
    &&\Sigma^{N,m,j' \rightarrow N+1,m,j}_{N,m,i' \rightarrow N+1,m,i}(s)
    \approx 
    \nonumber \\ 
    &&
    \Bigg\{
    \frac{\pi}{\hbar}\alpha_{\rm L} \Theta(\tilde E^{N}_{N+1}) 
    \Big[ 
         2\tilde E^{N}_{N+1} + (W^{N}_{N+1} - W^{N+1}_{N}) 
    \Big]
    \nonumber \\
    &&-
    \alpha_{\rm L} D(\tilde E^{N}_{N+1})
    \Big[
         2s + \frac{i}{\hbar}(W^{N}_{N+1} + W^{N+1}_{N})
    \Big] 
    \Bigg\}^{j'j}_{i'i} 
    \ ,
\end{eqnarray}
where  
$D(\tilde E^{N}_{N+1}) \approx 1+\gamma + 
\ln(\tilde E^{N}_{N+1}\delta/\hbar)$
and $\gamma \approx 0.58$ is Euler's constant.
The first term of (\ref{LEFT_RATE_LAPLACE}) is the standard 
Golden rule tunneling rate corresponding to the so-called 
orthodox theory of single-electron tunneling~\cite{Averin-Likharev}. 
The rate depends strongly on the charging energy difference, 
$\tilde E^{N}_{N+1}$, before and after the process,
which in the present problem is modified according to the quantum state 
of the qubit (the $W$ terms). At finite temperatures the step-function 
is replaced by $\Theta(E) \rightarrow  [1-\exp{(-E/k_{\rm B}T)}]^{-1}$.
We denote the full matrix of such rates $\hat \Gamma$ and we extensively 
exploit this matrix below. If the temperature is low and the applied 
transport voltage not too high, the leading tunneling process in the SET
is the sequential tunneling involving only two adjacent charge states, 
say $N=0$ and $N=1$.
We concentrate here on this case (to avoid confusion with 
the states of the qubit we will keep using the notation $N$ and $N+1$).

The second, logarithmically diverging part of 
(\ref{LEFT_RATE_LAPLACE}) produces a ``coherent-like'' term 
in the RHS of the master equation (\ref{MASTER_EQUATION}). 
These terms turn out to be unimportant in the first order of the 
perturbation theory. Indeed, for the left junction we obtain the 
following contribution to the RHS of (\ref{MASTER_EQUATION}):
\begin{equation}
\label{LEFT_DIVERGENT_TERM}
\alpha_{\rm L} \hat D_{\rm L}
\left(
{d\sigma\over dt} - 
{i\over \hbar}\left[\sigma,{\bar H}_{\rm qb}\right] 
\right)
\ ,
\end{equation}   
where ${\bar H}_{\rm qb}\equiv\frac{1}{2}(H_{\rm qb}(N)+H_{\rm qb}(N+1))$
and $\hat D_{\rm L}$ is a matrix in $N$ and $m$ spaces. The eigenvalues of the 
matrix $\hat D_{\rm L}$ are at most of order $D(\tilde E^{N}_{N+1})$.
Neglecting terms of order $\alpha_{\rm L} D(\tilde E^{N}_{N+1}) E_{\rm int}$ in 
Eq.(\ref{LEFT_DIVERGENT_TERM}), we can replace ${\bar H}_{\rm qb}$ by $H_0$. 
Our analysis shows that these neglected ``coherent-like''
terms do not change the results as long as 
$\alpha_{\rm L}\,|\ln(\tilde E^{N}_{N+1}\delta/\hbar)|\ll 1$.
Similar analysis can be made for the right tunnel junction 
of the SET. 

Now we can transfer all the ``coherent-like'' terms into the LHS of the master 
equation,
\begin{equation}
\label{MASTER_EQUATION_CORRECTED}
        \left(1-\alpha_{\rm L} \hat D_{\rm L}
          -\alpha_{\rm R} \hat D_{\rm R}\right) 
        \left\{
        {d\hat{\sigma}(t)\over dt} - {i\over\hbar}[\hat{\sigma}(t), H_0] 
        \right\}
        = \frac{1}{\hbar}\hat{\Gamma} \hat{\sigma}(t),
\end{equation}
and multiply Eq.~(\ref{MASTER_EQUATION_CORRECTED})
from the left by 
$(1-\alpha_{\rm L}\hat D_{\rm L} - 
\alpha_{\rm R}\hat D_{\rm R})^{-1} 
\approx (1 + \alpha_{\rm L}\hat D_{\rm L} + 
\alpha_{\rm R}\hat D_{\rm R})$ so that the
corrections move back to the RHS. 
Since $\hat \Gamma$ is itself linear in 
$\alpha_{\rm L},\alpha_{\rm R}$ the corrections 
belong to the second order in $\alpha$'s (more
accurately, they are small if 
$\alpha\,|\ln(\tilde E^{N}_{N+1}\delta/\hbar)| \ll 1$
for both junctions).
Thus, we drop the ``coherent'' corrections
and arrive at the final form of the master 
equation which we use below:
\begin{equation}
\label{MASTER_EQUATION_FINAL}
        {d\hat{\sigma}(t)\over dt} - {i\over\hbar}[\hat{\sigma}(t), H_0]  
        = \frac{1}{\hbar}\hat{\Gamma} \hat{\sigma}(t) \ .
\end{equation} 

To rewrite Eq.~(\ref{MASTER_EQUATION_FINAL}) in a matrix form
we have to choose a particular basis for the qubit. We choose the charge basis 
since in this basis the matrix $\hat\Gamma$ has the simplest form.
The matrix structure in the $m$ variable
simplifies considerably if we perform a Fourier transform with respect to $m$,
$\hat{\sigma}^{N}_{i,j}(k) \equiv \sum_m e^{ikm}\hat{\sigma}^{N,m}_{i,j}$.
To shorten formulas we introduce $A^N \equiv \hat{\sigma}_{0,0}^{N}(k)$,
$B^N \equiv \hat{\sigma}_{1,1}^{N}(k)$, 
$C^N \equiv \sum\limits_m e^{ikm}{\rm Re}\,\hat{\sigma}_{0,1}^{N,m}$,
and $D^N \equiv \sum\limits_m e^{ikm}{\rm Im}\,\hat{\sigma}_{0,1}^{N,m}$. 
Then, introducing a vector 
$X(k)=\big(A^{N},A^{N+1},B^{N},B^{N+1},C^{N},C^{N+1},D^{N},D^{N+1}\big)$,
we can rewrite (\ref{MASTER_EQUATION_FINAL}) as 
\begin{equation}
\label{ME8by8}
\dot X(k) = \frac{1}{\hbar}M(k)X(k) \ ,
\end{equation} 
where $M(k)$ is given by
{
\small
\arraycolsep=1pt
\begin{equation}
\label{MATRIX_8by8}
\left(
\begin{array}{ccccccccc}
-\Gamma_{L0}&e^{ik}\Gamma_{R0}&0&0&&0&0&E_{\rm J}&0\\
\Gamma_{L0}&-\Gamma_{R0}&0&0&&0&0&0&E_{\rm J}\\
0&0&-\Gamma_{L1}&e^{ik}\Gamma_{R1}&&0&0&-E_{\rm J}&0\\
0&0&\Gamma_{L1}&-\Gamma_{R1}&&0&0&0&-E_{\rm J}\\
\\
0&0&0&0&&-\Gamma_L&e^{ik}\Gamma_R&-E_{\rm ch}^{N}&0\\
0&0&0&0&&\Gamma_L&-\Gamma_R&0&-E_{\rm ch}^{N+1}\\
-\frac{1}{2}E_{\rm J}&0&\frac{1}{2}E_{\rm J}&0&&
E_{\rm ch}^{N}&0&-\Gamma_L&e^{ik}\Gamma_R\\
0&-\frac{1}{2}E_{\rm J}&0&\frac{1}{2}E_{\rm J}&&
0&E_{\rm ch}^{N+1}&\Gamma_L&-\Gamma_R
\end{array}
\right)
\end{equation}
}
and
\begin{eqnarray}
\label{ORIGINAL_RATES}
        \Gamma_L &\equiv& 2\pi\alpha_{\rm L} 
        [\; \mu_{\rm L} - E_{\rm SET}(1-2N_{\rm SET})] \; ,
\nonumber \\  
        \Gamma_R &\equiv& 2\pi\alpha_{\rm R}
        [-\mu_{\rm R} + E_{\rm SET}(1-2N_{\rm SET})]
\ ,
\end{eqnarray}
\begin{eqnarray}
\label{CORRECTED_RATES}
        \Gamma_{\rm L0} &\equiv& \Gamma_L + \Delta\Gamma_L \nonumber \; ,\\
        \Gamma_{\rm L1} &\equiv& \Gamma_L - \Delta\Gamma_L \nonumber \; ,\\ 
	\Gamma_{\rm R0} &\equiv& \Gamma_R + \Delta\Gamma_R \nonumber \; ,\\ 
	\Gamma_{\rm R1} &\equiv& \Gamma_R - \Delta\Gamma_R  \; .
\end{eqnarray}
The shifts $\Delta \Gamma$ are proportional to the interaction energy 
between the qubit and SET,
\begin{eqnarray}
\label{DELTA_GAMMA}
        \Delta\Gamma_{\rm L} &\equiv& 
	 \; \; \;    2\pi\alpha_{\rm L} E_{\rm int}  
        \nonumber \; ,\\
        \Delta\Gamma_{\rm R} &\equiv& 
	 - 2\pi\alpha_{\rm R} E_{\rm int}
        \ .
\end{eqnarray}
As we will see, these shifts are responsible for the separation of the peaks
of $P(m,t)$. Finally, the qubit's charging energies are given by
\begin{eqnarray}
\label{E_N_E_N+1}
&&E_{\rm ch}^{N} \equiv E_{\rm ch}(n_{\rm qb})
\ , 
\nonumber \\
&&E_{\rm ch}^{N+1} \equiv E_{\rm ch}(n_{\rm qb}) + 2E_{\rm int}
\ .
\end{eqnarray}

\subsection{Analysis of the master equation in the zeroth 
order in $E_{\rm J}$. Measurement and dephasing times.}

We are interested in the limit  
$E_{\rm J},E_{\rm int} \ll E_{\rm ch}(n_{\rm qb})$.
In this case
the system (\ref{ME8by8}) may be analyzed perturbatively
in $E_{\rm J}$. In zeroth order the system of equations (\ref{ME8by8})
factorizes into three independent groups. The first one,
\begin{eqnarray}
\label{A_SUBSYSTEM}
        \hbar\dot A^{N\phantom{+1}} &=&
        -\Gamma_{\rm L0} A^{N} + \Gamma_{\rm R0} e^{ik} A^{N+1}
\nonumber\; , \\
        \hbar\dot A^{N+1} &=& \phantom{-}\Gamma_{\rm L0} A^{N}  -
        \Gamma_{\rm R0} A^{N+1} \ ,
\end{eqnarray}
has exponential solutions $\propto e^{i\omega t}$ and eigenvalues
\begin{equation}
\label{A_EIGENVALUES}
        \hbar\omega_{1,2} = {i\over 2}(\Gamma_{\rm L0}+\Gamma_{\rm R0})
        \left\{1 \pm 
        \left[1+{4\Gamma_{0}(e^{ik}-1)
        \over \Gamma_{\rm L0}+\Gamma_{\rm R0}}\right]^{1\over 2}\right\} \ .
\end{equation}
Here the rate $\Gamma_0/\hbar$ of single-electron tunneling through 
the SET if the qubit is in the state $|0\rangle$ is given by
\begin{equation}
\label{GAMMA_ZERO}
\Gamma_{0} \equiv {\Gamma_{\rm L0}\Gamma_{\rm R0}
\over \Gamma_{\rm L0}+\Gamma_{\rm R0}}
\; .
\end{equation} 
When $k$ is small the first eigenvalue,
$\hbar\omega_1 \approx i(\Gamma_{\rm L0}+\Gamma_{\rm R0})$,
has a large imaginary part and the corresponding eigenmode quickly dies out.
The second eigenvalue $\hbar\omega_2 \approx \Gamma_{\rm 0}k + 
i\Gamma_{0} c_{0} k^2$ is small,
and the corresponding eigenmode, with
$A^{N+1}/A^{N} = \Gamma_{\rm L0}/\Gamma_{\rm R0}$,
survives. This solution describes
a wave packet propagating with the group velocity $\Gamma_{0}$,
which widens due to shot noise of the single electron
tunneling, its width being given by $\sqrt{\frac{c_{0}}{\hbar}\Gamma_{0}t}$.
The factor 
$c_{0} \equiv (\Gamma_{\rm L0}^2 + \Gamma_{\rm R0}^2)/
(\Gamma_{\rm L0} + \Gamma_{\rm R0})^2$ 
(see e.g. Ref.~\onlinecite{Korotkov_SET_Noise})
varies between $1/2$ in the 
symmetric situation ($\Gamma_{\rm L0} = \Gamma_{\rm R0}$) 
and $1$ in the extremely asymmetric case 
($\Gamma_{\rm L0}$ much larger or much smaller than $\Gamma_{\rm R0}$). 

Analogously the second group of equations for $B^N$ and $B^{N+1}$ 
describe a wave packet which moves in $m$-direction with the group velocity 
\begin{equation}
	\Gamma_{1} \equiv {\Gamma_{\rm L1}\Gamma_{\rm R1}
	\over \Gamma_{\rm L1}+\Gamma_{\rm R1}}
\end{equation}
and the width  growing as
$\sqrt{\frac{c_{1}}{\hbar}\Gamma_{1}t}$. The two peaks correspond
to the qubit in the states $|0\rangle$ and $|1\rangle$, respectively.
They separate when their distance is larger than their widths, i.e.\
$ \frac{1}{\hbar} |\Gamma_{0}-\Gamma_{1}|t \ge 
\sqrt{\frac{c_{0}}{\hbar}\Gamma_{0}t}
+\sqrt{\frac{c_{1}}{\hbar}\Gamma_{1}t}$.  
This means that after the time
\begin{equation}
\label{MEASUREMENT_TIME}
        t_{\rm ms} \equiv \hbar\left(
                          {\sqrt{\Gamma_{0}c_{0}}+\sqrt{\Gamma_{1}c_{1}}
                          \over  |\Gamma_{0}-\Gamma_{1}|}
                          \right)^2 \ , 
\end{equation} 
which we denote as the measurement time, the process
can constitute a quantum measurement.

To learn about the dephasing we analyze the last
four equations of (\ref{ME8by8}) at $k=0$,
which is equivalent to a trace over $m$.
These four equations may be recombined into two complex 
ones:
\begin{eqnarray}
\hbar{d\over dt}{\hat{\sigma}}_{0,1}^{N\phantom{+1}} 
&=& i E_{\rm ch}^{N\phantom{+1}}
\hat{\sigma}_{0,1}^{N\phantom{+1}}
-\Gamma_{\rm L}\hat{\sigma}_{0,1}^{N} + \Gamma_{\rm 
R}\hat{\sigma}_{0,1}^{N+1},
\nonumber \\
\hbar{d\over dt}{\hat{\sigma}}_{0,1}^{N+1} &=& i E_{\rm ch}^{N+1}
\hat{\sigma}_{0,1}^{N+1}
+\Gamma_{\rm L}\hat{\sigma}_{0,1}^{N} - \Gamma_{\rm 
R}\hat{\sigma}_{0,1}^{N+1} \ .
\label{CD_SUBSYSTEM}
\end{eqnarray}
The analysis of this set shows that if 
$|E_{\rm ch}^{N+1} - E_{\rm ch}^{N}| = 2E_{\rm int} 
\ll (\Gamma_{\rm L} + \Gamma_{\rm R})$
the imaginary parts of the eigenvalues are 
${\rm Im}\,\hbar\omega_1 \approx (\Gamma_{\rm L} + \Gamma_{\rm R})$ and 
${\rm Im}\,\hbar\omega_2 \approx {4E_{\rm int}^2\Gamma_{\rm L}\Gamma_{\rm R} / 
(\Gamma_{\rm L} + \Gamma_{\rm R})^{3}}$.
In the opposite limit 
$2E_{\rm int} \gg (\Gamma_{\rm L} + \Gamma_{\rm R})$ the imaginary parts are
${\rm Im}\,\hbar\omega_1 \approx \Gamma_{\rm L}$ and 
${\rm Im}\,\hbar\omega_2 \approx \Gamma_{\rm R}$. The first limit is physically
more relevant (we assume parameters in this regime),
although the second one is also possible if the tunneling
is very weak or the coupling between the qubit and the SET transistor is
strong. In both limits the dephasing time, which is defined as the
the longer of the two times,
\begin{equation}
\label{TAU_PHI_DEFINITION}
	\tau_{\phi} \equiv {\rm max}\{[{\rm Im}\,\omega_1]^{-1},
[{\rm Im}\,\omega_2]^{-1}\}
\end{equation}
is parametrically
different from the measurement time (\ref{MEASUREMENT_TIME}).
In the first limit, $2E_{\rm int} \ll (\Gamma_{\rm L} + \Gamma_{\rm R})$, it is
\begin{equation}
\label{DEPHASING_TIME}
     \tau_{\phi} = \hbar
     {(\Gamma_{\rm L} + \Gamma_{\rm R})^{3}\over 
     4 \Gamma_{\rm L} \Gamma_{\rm R} E_{\rm int}^2} 
\ .
\end{equation}
One can check that in the whole range of validity
of our approach the measurement time exceeds the dephasing time, 
$t_{\rm ms} > \tau_\phi$. This is consistent
with the fact that a ``good'' quantum measurement should completely
dephase a quantum state. 

The reason for the difference between the measurement time and the dephasing
time is the entanglement of the qubit's state with additional microscopic
states of the SET, --- states which cannot be characterized by the number of
electrons which have tunneled through the transistor, $m$, only.  Imagine that
we know (with high probability) that during some time from the beginning of the
measurement process exactly one electron has tunneled through the SET whatever
the qubit's state was.  Does it mean that with that high probability there was
no dephasing?  The answer is no.  The transport of electrons occurs via a real
state of the island, $N+1$.  The system may spend different times in this
intermediate state, i.e.  different phase shifts are acquired between the two
states of the qubit.  Since the time spent in the state $N+1$ is actually
random, some dephasing has occurred.  Next we can ask where the information
about this phase uncertainty is stored, i.e.  which states has the qubit become
entangled with.  These may not be the states with different $m$ or $N$ since 
$m$
is equal to one and $N$ is again equal to zero irrespective of the state of the
qubit.  The only possibility are the microscopic states of the middle island
and/or leads, which were subject to different time evolutions during different
histories of the tunneling process.  To put it in the language of
Ref.~\onlinecite{Stern_Aharonov_Imry}, the initial state of the system
$(a|0\rangle + b|1\rangle)\,|\chi\rangle\,|m=0\rangle$ evolves into
$a|0\rangle\,|\chi_0\rangle\,|m_0\rangle
+b|1\rangle\,|\chi_1\rangle\,|m_1\rangle$, where $|\chi\rangle$ stands for the
quantum state of the uncontrolled environment.  One may imagine a situation 
when
$m_0 = m_1$, but $|\chi_0\rangle$ and $|\chi_1\rangle$ are orthogonal.  In this
situation the dephasing has occurred but no measurement has been performed.

Dephasing was also analyzed in Refs.~\onlinecite{Levinson,Aleiner,Gurvitz}. 
In these works a quantum point contact (QPC) measuring device was used only 
as a source of dephasing, i.e.\ the information
on the current flowing in the measuring device was disregarded.  Thus a
distinction between measurement and dephasing was not made.  However the
expressions for the dephasing time were given by expressions similar to
(\ref{MEASUREMENT_TIME}).  As it became clear later
\cite{Korotkov}, the QPC does not involve an additional
uncontrolled environment during the measurement process and, therefore, $t_{\rm
ms} = \tau_\phi$.  Indeed, the tunneling in the QPC does not occur via a real
intermediate state and the discussion above does not apply.  In this sense the
QPC may be regarded as a $100\%$ efficient measuring device.  The additional
environment in the SET, which, as mentioned, reduces the efficiency of the
measuring device, plays, actually, a positive role in the quantum measurement,
provided it dephases the state of the qubit only when the system is driven out
of equilibrium.  This is because the quicker dephasing suppresses the
transitions between the states of the qubit (such a suppression of the
transitions due to a continuous observation of the quantum state of the system
is called the Zeno effect), while the initial probabilities
are preserved.  It is worth noting that a SET might also be used 
as a $100\%$ efficient device, if it was biased in the co-tunneling regime.  
However, this possibility has other drawbacks and hence is not considered 
here further.


\subsection{Higher orders in $E_{\rm J}$, the mixing time}
\label{Subsection_Mixing_Time}

Finally, we analyze what happens if we take into account the mixing 
terms proportional to $E_{\rm J}$ in the system (\ref{ME8by8}). 
We consider $k=0$ and investigate the eigenvalues of the $8\times8$ 
matrix (\ref{MATRIX_8by8}). Note that in the discussion above we have 
calculated all the eight eigenvalues for $E_{\rm J} = 0$ 
(the two eigenvalues of the complex system (\ref{CD_SUBSYSTEM}) are 
doubled when one considers it as a
system of four real equations). In the diagonal part there were two zeros,
which corresponded to two conserved quantities (for $k=0$), $A^{N}(0) +
A^{N+1}(0)=\hat{\sigma}_{0,0}$ and $B^{N}(0) + B^{N+1}(0)=\hat{\sigma}_{1,1}$.
Six other eigenvalues were large compared to the amplitudes of the mixing 
terms.
It is clear, that including the mixing, $E_{\rm J} \ne 0$, changes only 
slightly
the values of the six large eigenvalues. A more pronounced effect may be 
expected 
in the subspace of the two degenerate eigenvectors with zero eigenvalues.
It turns out that this degeneracy is lifted in second order of 
perturbation theory. One of the resulting eigenvalues remains zero.  
This corresponds to the conservation of the total trace
$A^{N}(0)+A^{N+1}(0)+B^{N}(0)+B^{N+1}(0)=1$.  The second eigenvalue acquires
now a small imaginary part and this gives the time scale of the mixing 
between the two states of the qubit. In the limit of (our) interest we obtain
\begin{equation}
\label{MIXING_TIME}
t_{\rm mix}^{-1} \approx  
	{4E_{\rm int}^2 E_{\rm J}^2 \Gamma 
        \over 
        \hbar \Delta E^2 
        (\Delta E^2 + (\Gamma_{\rm L}+\Gamma_{\rm R})^2)} 
\ , 
\end{equation}
where $\Gamma \approx \Gamma_{0} \approx \Gamma_{1}$, and
$\Delta E \approx \Delta E(\eta) \approx E_{\rm ch}^{N} 
\approx E_{\rm ch}^{N+1}$. Comparing Eq.~(\ref{MIXING_TIME})
and Eq.~(\ref{MEASUREMENT_TIME}) we see that both cases 
$t_{\rm ms} < t_{\rm mix}$ and $t_{\rm ms} > t_{\rm mix}$
are possible. 

Let us analyze a concrete physical situation.
We assume $\alpha_{\rm L} = \alpha_{\rm R} \equiv \alpha$.  
We choose $N_{\rm SET}$ far enough from the degeneracy point, 
which is $N_{\rm SET}=1/2$, so that $\Gamma_{\rm L} < \Gamma_{\rm R}$ 
and the related Coulomb blockade energy,
$E_{\rm CB} \equiv E_{\rm SET}(1-2N_{\rm SET})$, is of the order of $E_{\rm
set}$.  To satisfy the conditions for the Golden Rule 
(see (\ref{LEFT_RATE_LAPLACE}) and the discussion thereafter) we assume 
$E_{\rm CB}$ to be the largest energy scale of the system, 
$E_{\rm CB} \gg \Delta E$, and the chemical potential of
the left lead, $\mu_{\rm L}=V_{\rm tr}/2$, to exceed the Coulomb 
blockade energy by an amount of the order of $E_{\rm CB}$.  
The transport voltage should not, however, exceed the value 
beyond which further charge states of the SET transistor,
e.g.\ $N+2$ and $N-1$, become involved.  Thus $V_{\rm tr}/2 < E_{\rm
set}(1+2N_{\rm SET})$ and $N_{\rm SET}$ should be chosen far enough 
from zero as well. Thus we obtain
\begin{equation}
\label{MIXING_TIME_APPROX}
	t_{\rm mix}^{-1} 
	\approx 
        {2\pi\alpha E_{\rm SET} 
	E_{\rm int}^2 E_{\rm J}^2 
        \over 
        \hbar
        (\Delta E)^2 {\rm max}[(\Delta E)^2,(2\pi\alpha E_{\rm SET})^2]} 
        \ .
\end{equation}
The measurement time in the same regime is given approximately by
\begin{equation}
\label{MEASUREMENT_TIME_APPROX}
	t_{\rm ms}^{-1} \approx
	\frac{2\pi\alpha}{\hbar} {E_{\rm int}^2 \over E_{\rm SET}} \ .
\end{equation} 
For comparison we also give a rough value for the dephasing time,
which is short when  the measurement is performed. We assume the regime
in which Eq.~(\ref{DEPHASING_TIME}) is valid. Then
\begin{equation}
\label{DEPHASE_TIME_APPROX}
	{\tau_\phi}^{-1}  \approx  \frac{1}{2\pi\alpha \hbar}  {E_{\rm int}^2 
	\over E_{\rm SET}} \; .
\end{equation}
Note that the limit $\alpha \rightarrow 0$ is not allowed in 
(\ref{DEPHASE_TIME_APPROX}), since the validity of (\ref{DEPHASING_TIME})
eventually breaks down in this limit.

Thus, 
\begin{equation}
\label{MS_MIX_RATIO}
{t_{\rm ms}\over t_{\rm mix}}
\propto \left[{E_{\rm J} E_{\rm SET} \over \Delta E\, 
{\rm max}[\Delta E,2\pi\alpha E_{\rm SET}]}\right]^2
\ .
\end{equation} 
One recognizes two competing ratios here:  
$E_{\rm J} / \Delta E$, which is small, and 
$E_{\rm SET} / {\rm max}[\Delta E,2\pi\alpha E_{\rm SET}]$, which is large.  
The condition ${t_{\rm ms}/t_{\rm mix}} \ll 1$, thus,
imposes an additional constraint on the parameters of the system. 
On the other hand, for low conductance barriers in the SET
($\alpha < 1$ but not too small), the dephasing time is always 
shorter than the measurement time.

\subsection{Discussion and Extensions}

Let us summarize what has been done thus far:  To measure the quantum state of
the qubit we attach a SET to the qubit and consider the time evolution of the
whole system.  We provide two complimentary descriptions of the measurement
process.  In the first we trace over the SET's degrees of freedom and obtain a
reduced density matrix of the qubit.  From the time evolution of the last we
deduce the dephasing time and the mixing time, i.\ e.\ the relaxation times for
the off-diagonal and the diagonal elements of the reduced density matrix
respectively.  This description does not, however, contain information about 
the 
measurable quantity, the current in the SET.  Therefore, in a second 
description, we calculate the probability distribution, $P(m,t)$, 
of the number of electrons $m$ which have passed through the SET during time 
$t$.  
We will evaluate $P(m,t)$
numerically for parameters to be specified in the next section.  The results,
shown in Fig.~\ref{PLOT0.1} and ~\ref{PLOT0.3}, display the time evolution on
the time scale of $t_{\rm ms}$, relevant for the measurement process.

We observe that under appropriate conditions $P(m,t)$ splits into two separate
peaks, whose weights are given by the initial probabilities of the 
eigenstates of the qubit.  The splitting time is the minimum time after which 
one can distinguish between the states of the qubit and we call it, therefore, 
the
measurement time, $t_{\rm ms}$.  The measurement time turns out to be longer
than the dephasing time.  This fact indicates that the state of the qubit gets
entangled not only with the transport degree of freedom in the SET but also
with other, microscopic degrees of freedom which we do not
observe~\cite{Korotkov}.

Note, that the knowledge of $P(m,t)$ does not provide immediately the value of
the current flowing in the SET at all times.  Recently, attempts have been made
to describe the measurement process in terms of the current in the measuring
device \cite{Korotkov,Gurvitz_Measurement,Stodolsky}. 
However, further investigation of this problem is definitely needed.

\section{Discussion}
\label{Section_Discussion}

\subsection{Choice of parameters}
\label{Subsection_Choice_of_Parameters}

To demonstrate that the constraints on the circuit parameters can be met by
available technology, we summarize the constraints and suggest a suitable set.
The necessary conditions are:  $\Delta > E_{\rm qb} \gg E_{\rm J},E_L,k_{\rm
B}T$.  The temperature has to be chosen low to assure the initial
thermalization, $k_{\rm B}T\ll E_{\rm qb}$ and $k_{\rm B}T\ll \hbar
\omega_{LC}$, and to reduce the dephasing effects.  A good choice is $k_{\rm 
B}T
\sim E_{\rm J}/2$ since further cooling would not decrease the dephasing rate 
at
the degeneracy point much further.  Then, the dephasing rates during operations
(at the degeneracy point) and in the idle state (off degeneracy) are close to
each other.

Depending on the parameters chosen, the number of qubits $N$ and the total
computation time are restricted.  The dephasing times limits the number of
operations to $\tau_{\rm op} \ll \Gamma^{-1}_{\rm in},\Gamma^{-1}_\phi$.  This
is the only restriction for a single qubit.  If several qubits are coupled, the
dephasing for the whole system is faster.  Since the sources of dissipation for
different qubits are independent, the dephasing time gets $N$ times shorter.  
In addition, the energy splittings of the qubits should fit into the range 
provided by Eq.~(\ref{High_Frequency_Condition}), and the levels 
within this range should
be sufficiently different from each other to minimize the errors introduced by
non-zero inter-qubit coupling between 2-bit operations 
(cf. Section~\ref{Section_Qubits}). This provides a restriction on 
the number of qubits and the number of operations which can be performed. 
With circuit parameters discussed below and for not too many qubits 
these restrictions are weaker than those provided by the dephasing.
Moreover, we note that these restrictions are removed if the interaction is
effectively turned off.  This can be achieved on the ``software'' level, by the
use of refocusing voltage pulses.  On the ``hardware'' level this is achieved
with the improved design suggested in Ref.~\onlinecite{Makhlin}.  This allows
for larger numbers of qubits and longer computations.

As an example we suggest a system with  the following parameters:\\
(i) We choose junctions with the capacitance $C_{\rm J}=4\cdot 10^{-16}$F, 
corresponding to the charging energy (in temperature units) $E_{\rm qb} \sim 
2$K, and a smaller gate capacitance $C=2.5\cdot 10^{-18}$F to reduce the 
coupling to the environment. Thus at the working temperature of $T=50$mK 
the initial thermalization is assured. The superconducting gap has to be 
slightly larger  $\Delta > E_{\rm qb}$. We further choose $E_{\rm J}=100$mK, 
i.e.\ the time scale of one-qubit operations is $\tau_{\rm 
op}^{(1)}=\hbar/E_{\rm J}\sim 7\cdot 10^{-11}$s.\\ 
(ii) We assume that the resistor in the gate voltages circuit has $R\sim 
50\Omega$. Voltage fluctuations limit the dephasing time 
(\ref{Gamma_phi_Weiss}). We thus arrive at an estimate of the decoherence rate 
which allows for $(\Gamma_\phi \tau_{\rm op}^{(1)})^{-1}\sim 8\cdot 10^5$ 
coherent operations for a single bit.\\
(iii) To assure sufficiently fast 2-bit operations we choose $L\sim 3\mu$H. 
Then, the decoherence time is $(\Gamma_\phi \tau_{\rm op}^{(2)})^{-1}\sim 650$ 
times longer than a 2-bit operation.
In Table \ref{Tab_Alternative_Sets} we present alternative sets of parameters, 
assuming throughout that the resistance is fixed as $R=50\Omega$.
\parbox{11cm}{
\begin{table}
\narrowcaption
\caption{\label{Tab_Alternative_Sets}Examples of suitable sets of parameters.}
\vskip1mm
\begin{tabular}{ccccc|cc}
$C$     &$C_{\rm J}$    &$E_{\rm J}$    &$T$    &$L$&1-bit&2-bit\\
(aF)	&(aF)		&(mK)		&(mK)	&($\mu$H)&oper-s&oper-s\\
\hline
2.5     &400            &100            & 50    & 3 & $8\cdot10^5$&  650 \\
2.5	&400		&250		&125	& 1 & $8\cdot10^5$&  500 \\
2.5     &400            &250            &125    & 3 & $8\cdot10^5$& 1600 \\
40	&400		& 40		& 20	& 1 & $4\cdot10^3$&   85 \\
10	&400		&100		& 50	& 1 & $5\cdot10^4$&  200 \\
40	&400		&100		& 50	&0.5& $4\cdot10^3$&  100
\end{tabular}
\end{table}
}

\begin{figure}
\null\vspace{-2cm}
\epsfysize=27\baselineskip
\centerline{\hbox{\epsffile{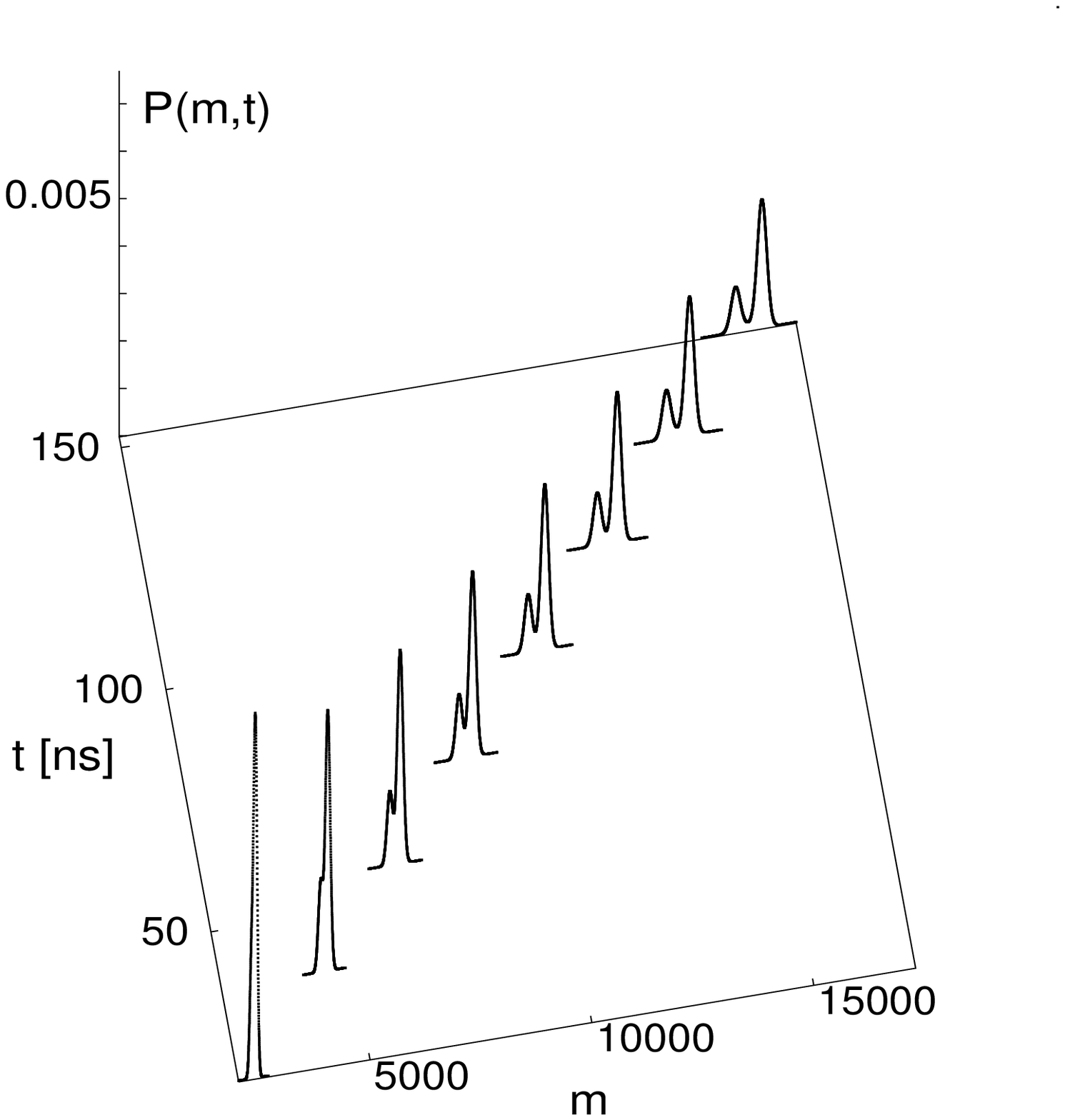}}}
\vspace{-5mm}
\narrowcaption
\caption[]{\label{PLOT0.1}
$P(m,t)$, the probability that $m$ electrons have tunneled during time
$t$. The parameters are those given in the text, $E_{\rm J} = 0.1$K. 
The time is measured in nanoseconds.
The initial amplitudes of the qubit's states: $a = \sqrt{0.75}$, 
$b=\sqrt{0.25}$.} 
\end{figure}

\begin{figure}
\null\vspace{-2cm}
\epsfysize=27\baselineskip
\centerline{\hbox{\epsffile{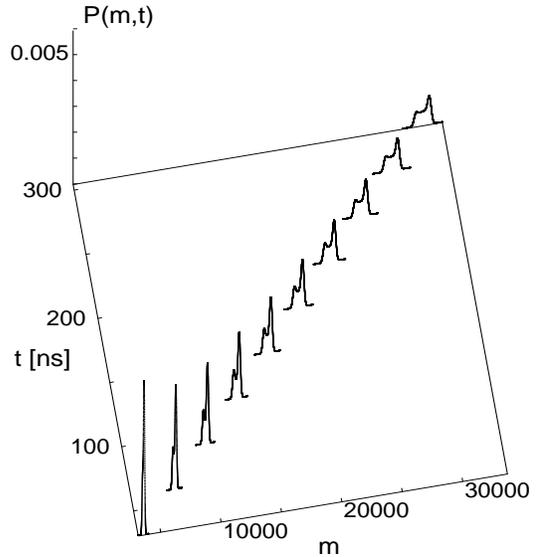}}}
\vspace{-5mm}
\narrowcaption
\caption[]{\label{PLOT0.3}
$P(m,t)$, the probability that $m$ electrons have tunneled during time
$t$. The parameters are those given in the text, $E_{\rm J} = 0.25$K.
The time is measured in nanoseconds.
The initial amplitudes of the qubit's states: $a = \sqrt{0.75}$,
$b=\sqrt{0.25}$.} 
\end{figure}

The quantum measurement process introduces additional constraints on the
parameters.  In order to demonstrate that the conditions assumed in this paper
are realistic we chose the charging energies $E_{\rm SET}$, $E_{\rm qb}$ and
$E_{\rm int}$ as follows:  The capacitance of the Josephson junction is $C_{\rm
J} = 4.0\cdot 10^{-16}$F, the gate capacitance of the qubit $C = 2.5\cdot
10^{-18}$F, the capacitances of the normal tunnel junctions of the SET\,
$C'=2.0\cdot 10^{-17}$F, the gate capacitance of the SET\, $C_{\rm g} = 
2.5\cdot
10^{-18}$F, and the capacitance between the SET and the qubit\, $C_{\rm int} =
2.5\cdot 10^{-18}$F.  We obtain:  $E_{\rm SET}\approx 20$K, $E_{\rm qb}\approx
2.5$K, $E_{\rm int}\approx 0.25$K.  Taking $n_{\rm qb}=0.35$, $N_{\rm SET} =
0.15$ and $eV_{\rm tr} = 48$K we get $\Delta E \approx 3$K, $E_{\rm CB} \equiv
E_{\rm SET}(1-2N_{\rm SET}) \approx 14$K, and $V_{\rm tr}/2 - E_{\rm CB} 
\approx
10$K (for definitions see subsection \ref{Subsection_Mixing_Time}).  We also
assume $2\pi\alpha = 0.1$.  The measurement time in this regime is $t_{\rm ms}
\approx 0.25 \cdot 10^{4} \hbar/(k_{\rm B}\,1{\rm K}) \approx 1.8 \cdot
10^{-8}$s.  For this choice of parameters we calculate $t_{\rm mix}$
using Eq.~(\ref{MIXING_TIME}). Assuming first $E_{\rm J} = 0.1$K 
we obtain $t_{\rm mix} \approx 1.4 \cdot 10^{5} \hbar/(k_{\rm B}\,1{\rm K}) 
\approx 1.0 \cdot 10^{-6}$s. Thus $t_{\rm mix}/t_{\rm ms} \approx 55$ and 
the separation of peaks should occur much earlier than the transitions 
happen.  Indeed, the numerical
simulation of the system (\ref{ME8by8}) for those parameters given
above shows almost ideal separation of peaks (see Fig.~\ref{PLOT0.1}).  On the
other hand, for $E_{\rm J} = 0.25$K, and we obtain $t_{\rm mix}/t_{\rm ms}
\approx 9$.  This is a marginal situation.  The numerical simulation in this
case (see Fig.~\ref{PLOT0.3}) shows that the peaks, first, start to separate,
but, later, the valley between the peaks is filled due to the mixing
transitions.

These numbers demonstrate that the quantum manipulations of Josephson junction
qubits, as discussed in this paper, can be tested experimentally using the
currently available lithographic and cryogenic techniques.  We have further
demonstrated that the current through a single-electron transistor can serve as
a measurement of the quantum state of the qubit, in the sense that in the case
of a superposition of two eigenstates it gives one or the other result with the
appropriate probabilities.

\subsection{Comparison with existing experiments}

The demonstration that the qubit is in a superposition of eigenstates should be
distinguished from another question, namely whether it is possible to verify
that an eigenstate of a qubit is actually a superposition of two different
charge states, which depends on the mixing angle $\eta$ as described by Eq.\
(\ref{n_OPERATOR}).  This question has been addressed in the experiments of
Ref.~\onlinecite{Bouchiat_PhD}.  They used a setup similar to the one shown in
Fig.~\ref{CIRCUIT}, a single-Cooper-pair box coupled to a single-electron
transistor.  They could demonstrate that the expectation value of the charge in
the box varies continuously as a function of the applied gate voltage as 
follows
from (\ref{n_OPERATOR}).  Another experiment along the same lines has been
performed by Nakamura et al.  \cite{Nakamura}.  They demonstrated by
spectroscopy that the energy of the ground state and of the first excited state
have the expected gate-voltage dependence of superpositions of charge states.

Our theory can also describe stationary-state measurements of
Ref.~\onlinecite{Bouchiat_PhD}.  The measurement was performed in the
whole range of voltages including the vicinity of the degeneracy point, and we
need a different pertrubation expansion which is valid in this
region~\cite{Shnirman_PRB}.  For this purpose we have to analyze the rates in
the master equation (\ref{MASTER_EQUATION_FINAL}) for general values of the
mixing angle $\eta$, relaxing the requirement $\eta\ll 1$.  To this end we
rewrite the master equation (\ref{MATRIX_8by8}) in the qubit's eigenbasis
(\ref{Eigen_Basis}) and consider the mixing terms as a perturbation.  Then, the
expansion parameter is $E_{\rm J}E_{\rm int}/\Delta E^2(\eta)$ which extends 
the
region of the validity of the perturbative treatment.
It turns out that each eigenstate of the qubit,
$|0\rangle$ or $|1\rangle$,  corresponds to a single,
though $\eta$-dependent tunneling rate $\Gamma_{0/1}(\eta)$.
Thus, if the qubit is prepared in one 
of its eigenstates, only one peak is observed.
This is the case even close to the degeneracy point 
$\eta\approx\pi/2$ where the eigenstates are
superpositions of two charge states with substantial weight of both components.
On the other hand a charge state, being a superposition of two eigenstates, 
would produce two peaks in the current distribution. In this sense the 
measurement provides information about occupation probabilities of the 
eigenstates rather than charge states of the qubit.  The correction to the 
tunneling rate $\Gamma_{0/1}(\eta)-\Gamma\propto\pm\cos^2\eta$ is proportional 
to the average charge in the corresponding eigenstate of the qubit, in 
accordance with the experiments of Bouchiat et al.~\cite{Bouchiat_PhD}.
Close to the degeneracy it is more difficult (takes longer) to distinguish the 
eigenstates since the peaks get closer.

In our solution of the master equation we neglected dissipative effects due to
the environment discussed in Section~\ref{Section_Dephasing}.  It is justified
on time scales shorter than the environment-induced relaxation and dephasing
times given by (\ref{Gamma_r_Weiss}), (\ref{Gamma_phi_Weiss}).  It is also
justified at longer times as long as the environment-induced effects are weaker 
than the measurement-induced mixing, i.e. the rates (\ref{Gamma_r_Weiss}), 
(\ref{Gamma_phi_Weiss}) are smaller than the mixing rate
(\ref{MIXING_TIME}).  In this limit thermal relaxation is ineffective, and
in the stationary regime ($t\to\infty$)
the qubit is in the equally-weighted mixture of two states, corresponding to 
the 
infinite effective temperature. The current in the SET,
$e\Gamma$, is insensitive to the gate voltage in the qubit's circuit.  In the
opposite limit of weaker mixing, which is relevant to the experiments of
Bouchiat~et~al.~\cite{Bouchiat_PhD}, mixing is ineffective and thermal
relaxation takes over.  At low temperature $k_{\rm B}T\ll E_{\rm J}$ 
dissipation
keeps the system in the ground state, and the stationary current value is
$\frac{e}{\hbar}\Gamma_0(\eta)$.

\subsection{Related theories}

It is also interesting to compare our proposal with
the ``quantum jumps'' technique employed in quantum
optics in general, and with the realizations of the
qubits by trapped ions in particular (for a review see
Ref.~\onlinecite{Quantum_Jump}).
Indeed, the concepts are very close in spirit: the state of
the system is examined by an external nonequilibrium current
(electrons in our case and photons in the quantum jumps
technique). There is, however, an important difference. 
In the quantum jumps measurements only one of
the logical states  scatters photons. Therefore,
the efficiency of the measurement is limited  
by the ability to detect photons. In principle we
could realize this situation also in our system, 
if we bias the SET transistor such that different states of the
qubit switch the transistor between the off and on regimes.
Then the efficiency of the measurement is determined by the ability
to detect individual electron --- which is possible
in single-electron devices, for instance by charging a
single-electron box --- and the measurement time would be
given by the time it takes the first electron to tunnel.
However, this mode of operation would require that the SET 
transistor is kept near the switching point, where thermal 
fluctuations and higher order processes could modify the picture 
substantially. Therefore, we have concentrated here on a 
situation in which the SET transistor conducts for both
states of the qubit, and the measurement
requires distinguishing large numbers of charges or macroscopic currents.
Accordingly, the measurement time is limited by the shot noise.

\subsection{Summary}

To conclude, the fabrication of Josephson junction
qubits is possible with current technology. In these  systems 
fundamental features of macroscopic quantum-mechanical systems can be 
explored. More elaborate designs  as  
well as further progress of nano-technology, will provide longer
coherence times  and allow scaling to larger numbers of qubits. 
The application of Josephson junction systems as
elements of a quantum computer, i.e.\ with a very large number of manipulations
and large number of qubits, will remain a challenging issue, 
demanding in addition to the perfect control of time-dependent gate voltages
a still longer phase coherence time. We stress, however, that many aspects
of quantum information processing can initially be tested on 
simple circuits as proposed here.

We have further shown that a single-electron transistor
capacitively coupled to a qubit may serve as a quantum measuring device
in an accessible range of parameters. We have described the process of 
measurement by deriving the time evolution of the reduced density matrix. 
We found that the dephasing time is shorter than the measurement 
time, and we have estimated the mixing time, i.\,e. 
the time scale on which the transitions induced by the measurement occur.


\acknowledgments
We thank E.~Ben-Jacob, T. Beth,
C.~Bruder,  L.~Dreher, Y.~Gefen, S.~Gurvitz, Z.~Hermon, J. K\"onig, 
A.~Korotkov,
Y.~Levinson, J.~E.~Mooij, T.~Pohjola, and H.~Schoeller  for stimulating
discussions. This work has been supported by a Graduiertenkolleg 
 and the SFB 195 of the DFG, the A.\ v.\ Humboldt foundation (Y.M.)
and the German Israeli Foundation (Contract G-464-247.07/95) (A.S.).

\end{multicols}


\begin{thebibliography}{10}

\bibitem{Likharev}
K.K.~Likharev. Proc. IEEE {\bf 87}, 606 (1999).

\bibitem{Barenco_Review}
S.~Lloyd, Science {\bf 261}, 1589 (1993);
C.H.~Bennett, Physics Today {\bf 48} (10), 24 (1995);
A.~Barenco, Contemp. Phys. {\bf 37}, 375 (1996);
D.P.~DiVincenzo, Science {\bf 269}, 255 (1995),
and in {\it Mesoscopic Electron Transport}, ed. L.L.~Sohn et al., Kluwer, 1997.

\bibitem{Zoller}
J.I.~Cirac and P.~Zoller, Phys. Rev. Lett. {\bf 74}, 4091 (1995).

\bibitem{Wineland}
B.E.~King, C.S.~Wood, C.J.~Myatt, Q.A.~Turchette, D.~Leibfried, W.M.~Itano,
C.~Monroe, and D.J.~Wineland, Phys. Rev. Lett. {\bf 81}, 1525 (1998).

\bibitem{Loss}
B.E.~Kane, Nature {\bf 393}, 133 (1998);
D.~Loss and D.P.~DiVincenzo, Phys. Rev. A {\bf 57}, 120 (1998);
G. Burkard {\sl et al.}, Phys. Rev. B {\bf 59}, 2070 (1999).

\bibitem{Lukens}
R.~Rouse, S.~Han, and J.E.~Lukens, Phys. Rev. Lett. {\bf 75}, 1614 (1995).

\bibitem{Mooij}
J.E.~Mooij, T.P~Orlando, L.~Levitiov, Lin Tian, C.~H.~van~der~Wal,
and S. Lloyd, Science {\bf 285}, 1036 (1999).

\bibitem{Ioffe} 
L.B.~Ioffe, V.B.~Geshkenbein, M.V.~Feigel'man,
A.L.~Fauchere, and G.~Blatter, 
Nature {\bf 398}, 679 (1999).

\bibitem{Rome}
C.~Cosmelli {\sl et al.}, to be published in
`Quantum Coherence and Decoherence -- ISQM -- Tokyo 98', 
North-Holland Publ. Delta Series.

\bibitem{Our_PRL}
A.~Shnirman, G.~Sch\"on, and Z.~Hermon, Phys. Rev. Lett. {\bf 79}, 2371 (1997).

\bibitem{Bouchiat_PhD}
V.~Bouchiat, Ph.~D. Thesis, Universit\'e Paris 6 (1997);
V.~Bouchiat, D.~Vion, P.~Joyez, D.~Esteve, and M.H.~Devoret,
Physica Scripta {\bf T76}, 165 (1998).

\bibitem{Geerligs}
A.~Maassen van den Brink, G.~Sch\"on, and L.J.~Geerligs, Phys. Rev. Lett. {\bf
67}, 3030 (1991).

\bibitem{Parity}
M.T.~Tuominen, J.M.~Hergenrother, T.S.~Tighe, and M.~Tinkham,
Phys.~Rev.~Lett. {\bf 69}, 1997 (1992).

\bibitem{Siewert}
J.~Siewert and G.~Sch\"on, Phys. Rev. B {\bf 54}, 7421 (1996).

\bibitem{Nakamura}
Y.~Nakamura, C.D.~Chen, and J.S.~Tsai, Phys. Rev. Lett. {\bf 79}, 2328 (1997).

\bibitem{Hadley}
P.~Hadley, E.~Delvigne, E.H.~Visscher, S.~L\"ahteenm\"aki, and J.E.~Mooij,
Phys.  Rev. B {\bf 58}, 15317-20 (1998).

\bibitem{Nakamura_Nature}
Y.~Nakamura, Yu.A.~Pashkin, and J.S.~Tsai,  
Nature {\bf 398}, 786 (1999).

\bibitem{Averin}
D.V.~Averin, Solid State Commun. {\bf 105}, 659 (1998).

\bibitem{Makhlin}
Yu.~Makhlin, G.~Sch\"on, and A.~Shnirman, Nature {\bf 386}, 305 (1999).

\bibitem{Shnirman_PRB}
A.~Shnirman and G.~Sch\"on, Phys. Rev. B {\bf 57}, 15400 (1998).

\bibitem{Levinson}
Y.~Levinson, Europhys. Lett. {\bf 39}, 299 (1997).

\bibitem{Aleiner}
I.L.~Aleiner, N.S.~Wingreen, Y.~Meir, Phys. Rev. Lett. {\bf 79}, 3740 (1997).

\bibitem{Gurvitz}
S.A.~Gurvitz, Phys. Rev. B {\bf 56}, 15215 (1997).

\bibitem{Buks}
E.~Buks, R.~Schuster, M.~Heiblum, D.~Mahalu, and V.~Umansky, Nature, {\bf 391},
871 (1998).

\bibitem{Lafarge}
P.~Lafarge, P.~Joyez, D.~Esteve, C.~Urbina, and M.H.~Devoret, Phys. Rev. Lett. 
{\bf 70}, 994 (1993).

\bibitem{Schoen-Zaikin94}
G.~Sch\"on and A.D.~Zaikin, Europhys.\ Lett. {\bf 26}, 695 (1994).

\bibitem{Tinkham2}
M.~Tinkham, {\it Introduction to Superconductivity}, 2nd edition, McGraw-Hill
(1996).

\bibitem{Caldeira-Leggett}
A.O.~Caldeira and A.J.~Leggett, Ann. Phys. (NY) {\bf 149}, 374 (1983).

\bibitem{Two_Level_Leggett}
A.J.Leggett, S.~Chakravarty, A.T.~Dorsey, 
M.P.A.~Fisher, A.~Garg, and 
W.~Zwerger, Rev. Mod. Phys. {\bf 59}, 1 (1987).

\bibitem{Two_Level_Weiss}
U.~Weiss and M.~Wollensak, Phys. Rev. Lett. {\bf 62}, 1663 (1989);
R.~G\"orlich, M.~Sassetti and U.~Weiss, Europhys. Lett. {\bf 10}, 507 (1989).

\bibitem{P(E)_Panyukov_Zaikin}
S.V.~Panyukov and A.D.~Zaikin, J. Low. Temp. Phys. {\bf 73}, 1 (1988).

\bibitem{P(E)_Odintsov}
A.A.~Odintsov, Sov. Phys. JETP {\bf 67}, 1265 (1988).

\bibitem{P(E)_Nazarov}
Yu.V.~Nazarov, Sov. Phys. JETP {\bf 68}, 561 (1989).

\bibitem{P(E)_Devoret}
M.H.~Devoret, D.~Esteve, H.~Grabert, G.L.~Ingold, and H.~Pothier, Phys. Rev.
Lett. {\bf 64}, 1824 (1990).

\bibitem{Hakim_Ambegaokar}
V.~Hakim and V.~Ambegaokar, Phys. Rev. B {\bf 32}, 423 (1985).

\bibitem{Golubev_Zaikin_Dephasing}
D.S.~Golubev and A.D.~Zaikin,
In {\it Quantum Physics at Mesoscopic Scale},
eds. T.~Glattli, M.~Sanquer and Tran Thanh Van,
Frontieres, Gif-sur-Yvette, 1999; cond-mat/9907497.

\bibitem{Kouwenhoven}
L.P.~Kouwenhoven, private communication.

\bibitem{errorcorrectingcodes}
P.W.~Shor, Phys. Rev. A, {\bf 52}, 2493 (1995);
A.~Steane, Proc. Roy. Soc. of London A, {\bf 452}, 2551 (1996);
E.H.~Knill and R.~Laflamme   Phys. Rev. A {\bf 55}, 900, (1997);
E.~Knill, R.~Laflamme and W.~Zurek, Science, {\bf 279}, 342 (1998).

\bibitem{Schon_Zaikin_Review}
V.~Ambegaokar, U.~Eckern and G.~Sch\"on,
Phys. Rev. Lett. {\bf 48}, 1745 (1982);
G.~Sch\"on and A.D.~Zaikin, Physics Reports {\bf 198}, 237 (1990).

\bibitem{Schoeller_PRB}
H.~Schoeller and G.~Sch\"on, Phys. Rev. B {\bf 50}, 18436 (1994).

\bibitem{Nazarov}
Yu.V.~Nazarov, Physica B {\bf 189}, 57 (1993);
T.H.~Stoof and Yu.V.~Nazarov, Phys. Rev. B {\bf 55}, 1050 (1996);
S.A.~Gurvitz and Ya.S.~Prager, Phys. Rev. B {\bf 53}, 15932 (1996).

\bibitem{Averin-Likharev}
D.V.~Averin and K.K.~Likharev, in {\it Mesoscopic Phenomena in Solids},
edited by B.L.~Altshuler, P.A.~Lee, and R.A.~Webb (Elsevier, Amsterdam,
1991), p. 173.

\bibitem{Korotkov_SET_Noise}
A.~Korotkov, Phys. Rev. B {\bf 49}, 10384 (1994).

\bibitem{Stern_Aharonov_Imry}
A.~Stern, Y.~Aharonov, and Y.~Imry, Phys. Rev. A {\bf 41}, 3436 (1990).

\bibitem{Korotkov} 
A.N.~Korotkov, Phys. Rev. B {\bf 60}, 5737 (1999).

\bibitem{Gurvitz_Measurement}
S.A.~Gurvitz, preprint, quant-ph/9808058.

\bibitem{Stodolsky}
L.~Stodolsky, Phys. Lett. B {\bf 459}, 193 (1999).

\bibitem{Quantum_Jump}
D.J.~Wineland, C.~Monroe, W.M.~Itano, D.~Leibfried, B.~King, and
D.M.~Meekhof, J. Res. Natl. Inst. Stand. Tech. {\bf 103}, 259 (1998).

\end{thebibliography}
\end{document}